\documentclass[preprint]{aastex}

\newcommand{\subsun}{\mbox{$_{\odot}$}}

\newcommand{\etal}{{\it et al.\/}}

\newcommand{\irfilt}{$J$, $H$ and $K_s$}
\newcommand{\SB}{mag DN$^{-1}$ arcsec$^{-2}$}

\newcommand{\ntotal}{105}
\newcommand{\nvdirect}{61}
\newcommand{\nirdirect}{104}

\begin{document}

\title{Integrated Light 2MASS IR Photometry of Galactic Globular Clusters}

\author{Judith G. Cohen\altaffilmark{2}, Scott Hsieh\altaffilmark{2},
Stanimir Metchev\altaffilmark{3}, S. G.~ Djorgovski\altaffilmark{2}
\& M. Malkan\altaffilmark{3}}

\altaffiltext{2}{Palomar Observatory, Mail Stop 105-24,
California Institute of Technology, Pasadena, Ca., 91125, 
jlc(george)@astro.caltech.edu, scotthsieh@gmail.com}

\altaffiltext{3}{Department of Physics and Astronomy, University
of California, Los Angeles, Ca. 90095, metchev(malkan)@astro.ucla.edu}

\begin{abstract}

We have mosaiced 2MASS images  to derive surface
brightness profiles  in \irfilt\ for 104 Galactic
globular clusters.  We fit these with King profiles, and show
that the core radii are identical to within the errors
for each of these IR colors, and are identical to the core
radii at $V$ in essentially all cases.  We derive integrated
light colors $V-J, V-H, V-K_s, J-H$ and $J-K_s$ for these
globular clusters.  Each color shows a reasonably tight relation
between the dereddened colors and metallicity.  Fits to these
are given for each color.
The IR--IR colors have very small errors due largely to
the all-sky photometric calibration of the 2MASS survey,
while the $V-$IR colors have substantially larger uncertainties.
We find fairly good agreement with measurements of integrated light
colors for a smaller sample of Galactic globular clusters
by Aaronson, Malkan \& Kleinmann from 1977. Our results
provide a calibration for the integrated light of distant
single burst old stellar populations from very low to Solar
metallicities.  A comparison of our 
dereddened measured colors
with predictions from several
models of the integrated light of single burst old populations
shows good agreement in the low metallicity domain for $V-K_s$ colors, but
an offset at a fixed [Fe/H] of $\sim$0.1 mag in $J-K_s$, which
we ascribe to photometric system transformation issues.
Some of the models fail to reproduce the behavior of the integrated
light colors of the Galactic globular clusters 
near Solar metallicity.  

\end{abstract}


\section{Introduction}

Galactic globular clusters (GCs) are of great interest as nearby
representatives of simple stellar systems all of whose
stars share the same age and initial chemical 
composition\footnote{We ignore the anomalous GC $\omega$ Cen here.}.
They are close enough that individual stars can be studied in
detail with spectroscopy and photometry, while far enough
away that, with some difficulty, their integrated light can
be measured as well. Their ages and initial mass functions can be determined
through analysis of deep high spatial resolution imaging, primarily
from HST; mass
segregation can be studied for these objects as well. 
Integrated light measurements of Galactic GCs are of key importance
as they provide calibration data for the study of more distant
early type galaxies with predominantly old populations as well
as the GCs of distant galaxies for which only the
integrated light can be observed. 

The Two Micron All Sky Survey (2MASS) affords us a wonderful 
opportunity to study 
the surface brightness profiles of nearby GCs  in the 
infrared.  As described in detail by
\cite{2mass1}, two 1.3m diameter telescopes were used, one in the
northern and one in the southern hemisphere.
The whole sky was observed 
in three colors, $J$, $H$, and $K_s$ (a variant of the $K$
filter described in detail therein).  The advantages
for our purposes of the 2MASS
data over any previously existing are many.  The photometry is all-sky
with careful attention to calibration issues, ensuring uniformity over
all the frames.  The database is digital, and hence background
subtraction and sophisticated image analyses are feasible.
There are, of course, disadvantages as well.
The effective exposure time for each point in the sky in this survey was
short, only 7.8 sec in each of the three colors, so these images
reach a relatively shallow limiting magnitude.
Furthermore the spatial resolution is limited by
the adopted detector pixel size of 2.0'' on a side.

\section{The Surface Brightness Profiles from 2MASS}

We describe here the process we have used to define and fit the surface brightness
profile of Galactic GCs in \irfilt\ from the 2MASS data.  
The integrated light in the IR is
sensitive to rare bright RGB tip giants and AGB stars.  Hence
centroiding in the IR 
is  more subject to stochastic effects dependent on cluster richness
than it is at optical wavelengths, where the numerous stars near the main sequence
turnoff make a substantial contribution to the integrated light.
We therefore begin by adopting the GC cluster centers 
determined from the  on-line
database for the Galactic GC 
system maintained by W. Harris 
\citep{harris96}\footnote{at 
http://physwww.mcmaster.ca/${\sim}$Eharris/WEHarris.html}, 
as updated in 2003 (henceforth H96).  These were derived from optical images of 
the GCs.  The tidal radii of GCs are determined by the gravitational field
of the Galaxy.
Furthermore, there is no hope of accurately measuring the surface brightness
profile in the outermost parts
of the Galactic GCs from the short and relatively shallow 2MASS exposures
as the GCs are often very extended and
the surface brightness near the tidal radius ($r_t$) is low.  
We thus adopt the values of $r_t$ determined from optical photometry
as compiled in the same on-line database; these are primarily from 
\cite{trager95} (henceforth TKD95).

A square region, centered on the GC and with a side 
length of 2$r_t$, was used for each cluster.
We downloaded the FITS image files from
the 2MASS website Batch Image service that cover the 
required area for each Galactic GC for each of the three filters
\irfilt.  The most extended GCs
(clusters with $r_t \gtrsim 0.45^{\circ}$) are likely 
to have some missing data (not a problem), and 
those with $r_t \gtrsim 0.6^{\circ}$ (${\sim}5$ Galactic GCs)
could not be analyzed at all due to excessive memory and CPU requirements.
Data for one such very extended GC, 47 Tuc, was recovered from the 
2MASS Large Galaxy Catalog.

Processed frames from 2MASS have 1.0'' x 1.0'' pixels and are
tiled together in a regular 
fashion, with an overlap region between frames. The individual frames 
were made into 
a mosaic for each GC in each of the three IR colors 
by examining the header in each FITS file to determine if it had 
neighboring 
frames adjacent to it, and then removing duplicated points in the overlap 
regions. 
Because the tiling was not perfectly gridlike but rather somewhat staggered,
a few duplicate points were left behind and  a few other points
were removed that  should have been kept,
but this should have a negligible effect. When the frames were irregularly stacked 
(as was the case in certain clusters), the mosaic procedure failed. 
This was fixed when it was not too difficult to do so by deleting selected frames 
from the data set; less than 5 GCs were handled this way.

Frames were adjusted to a constant sky value and calibrated to a universal 
adopted
magnitude zero point for each color; we adopted
values typical of those on the 2MASS images, specifically 20.45, 20.90, and 19.93 
\SB\ for \irfilt,
respectively.  The calculation relies on the photometric zero point
determined for each frame by the 2MASS project which is given in
the keyword MAGZP in the header of each image.
Extinction within
the Earth's atmosphere is included in these zero points. 
The value of each pixel is reset to reflect the different depth achieved
by each particular frame in the mosaic for each GC in each of the three colors.
The sky value for each GC at each of \irfilt\ was simply taken to be the 
mean sky value of 
the individual rescaled frames, as indicated by the keyword SKY in the
FITS header of each file.  A detailed discussion of the algorithms used by
the 2MASS project to determine the values of these parameters
for each frame is given in \cite{2mass2}.  The background from non-member
stars in the region of each GC was evaluated in a region 
largely beyond
the tidal radius, extending from 0.95 to 2.0 $r_t$.

The calculation of the surface brightness followed that of 
\cite{fischer92}, based on the method of \cite{djorgovski86}. 
The image was divided 
into several annuli centered on the globular cluster.
The 2MASS detector pixels are  2.0'' on a side, so the 
first annulus 
was a circle  with a 5'' radius. Subsequent annuli went 
from 5'' to 95\% of the tidal radius, with the annuli evenly spaced 
in logarithmic space. If $r_t < 5$', 8 annuli 
were used; if less 
than 10', 10 annuli were used; if greater than 10', 15 
annuli were used. 
The effective radius, $r_{eff}$, assigned to each annulus is the intensity-weighted
mean radius of the annulus (under the assumption of a linear intensity profile
over the width of the annulus; Newell \& O'Neil 1978),

$$ r_{eff} = {{2r_2^3 - r_1^3}\over{3r_2^2 - r_1^2}}, $$

\noindent where $r_1$ and $r_2$ are the inner and outer radii of the annulus.

Each annulus was divided into eight 45$^{\circ}$ sectors,
and the mean DN of the pixels within each sector was found. 
In the first annulus (actually circle), as well as all annuli
with $r_{eff} < 10$'', the mean  
of the 8 sector means was used as the surface brightness within
that annulus, since the variations among the sectors for a given $r_{eff}$  
arise primarily from possible stochastic fluctuations and potential
errors in the adopted position of the cluster center rather than from
photometric errors or from non-member stars. For annuli
with $r_{eff} > 10$'', the median of these 8 sector means
(in practice the average of the fourth and fifth value of the sorted list)
was chosen as 
the final value for that annulus. 
Use of the median
minimizes the surface brightness fluctuations due to a small number of
bright stars, but biases the mean surface brightness towards lower values,
since it effectively excludes some light from the brightest red giants;
however, this surface brightness profile is more representative of the
bulk of the stellar population in the cluster.

\subsection{Globular Cluster Centers}

We initially adopt the centers of the GCs from the
2003 on line version of the database of \cite{harris96}.
The majority of the cluster center coordinates in 
both H96 and in \cite{djorg93}
are from \cite{shawl86}, and for most cases they are not expected
to be determined better than to a few arcsec.
While the centers from H96
generally appeared to be correct, sometimes 
they seemed not to coincide with the  
center of a few of the GCs 
as judged from the 2MASS images. An effort was made to develop a centroiding
routine that would operate on the 2MASS images, but stochastic effects made
this difficult.  Instead, a specific list, based on visual inspection
of the 2MASS images, was made of GCs which might have centers in error
by more than 2 arcsec from their nominal values. 
Only those GCs for which
the entire distribution of IR light appears shifted with respect
to the nominal optical center are listed.
GCs with an asymmetric distribution of the brightest giants within 1 core radius
around the optical center such as NGC~5927 and NGC~6712 are not included here;

The art of determining the location of the center of light for
a GC is not trivial.  We will demonstrate later that small 
errors of $\sim$4 arcsec in cluster
center position are probably common in the H96 database, as was also
found by \cite{noyola06} who analyzed archival HST images of the
central regions of 38 Galactic GCs.

The new positions we thought might be appropriate for the centroids
of eight GCs
were carried along as additional clusters in the analysis and are listed in
Table~\ref{table_center}; a few additional cases of apparent
centroid error are given in the last part of the table.
For example, NGC~6541 has a central position in the 2003 version of 
the database of \cite{harris96} which  is more than 12' away from 
its true location.   The differences in $K_s$ and in $J-K_s$
for these small shifts for the 8 GCs
are given in the fifth and sixth columns of this table. The resulting
change in total brightness for these
small differences in adopted GC central position
can reach $\sim$0.15 mag, which would
directly affect an optical-IR color such as $V-K$.   
The shift in a 2MASS-2MASS color such as $J-K_s$ is smaller.  The 
position for the cluster center adopted
here for the 8 GCs listed in the first part of this table
is that which gave the larger signal at \irfilt.

\subsection{Eliminating Bright Field Stars}

The field star background is major concern, especially for the GCs
seen against the Galactic bulge.  While one could produce a  CMD
for the entire field of the mosaic image of a GC
and then eliminate stars that do not lie along the expected cluster
isochrone, we chose to adopt a scheme which 
is much easier to implement yet still succeeds in eliminating
most of the brightest field stars.   We
calculate the $K$ magnitude of the RGB tip for each cluster from its
known distance and interstellar reddening assuming $M_K(tip) = -5.9$ mag.  
We then add a buffer of 1.5 mag\footnote{If we were doing this again,
we would use a smaller value for the buffer.}.  Stars brighter than
this $K$ are too bright to be cluster members.  They
were identified in the field of each GC from the 2MASS Point Source 
Catalog. The $K_s$ band was used to select non-members; point sources 
that were deemed 
non-members in the $K_s$ band were also removed in the J and H bands.
A 5 x 5 pixel area around such stars was deleted from the mosaiced image
for sources fainter 
than $K_s = 10$, while for brighter stars, an area 11 x 11 pixels was deleted. 
This cleaning operation could not be carried out
close to the core of the GC 
where there might be crowding, which region is larger than expected
due to the low spatial resolution of 2MASS, so we only selected and 
eliminated such sources 
for $r > 0.3r_t$. 

The uncertainty in the surface brightness for each $r_{eff}$ was 
the rms dispersion of the 8 sector values divided by two, taken
in quadrature with the dispersion of the background measurements from
the set of frames for a given cluster.  The factor of two 
instead of ${\sqrt8}$ reflects the difference between 
$\sigma$ around a median instead
of a mean \citep[see, e.g.][]{lupton93}.

\subsection{Fitting the IR Surface Brightness Profiles}

For each GC, the empirical King profile \citep{king62} is fit to the 
surface brightness profile we determined for each of the colors \irfilt. 
We note that this
formula is distinct from the dynamical King models \citep{king66a}, which
were used in deriving the structural parameters in TKD95 and H96;
however, for the purposes of the present paper, the effective
differences are expected to be sufficiently small in the radial range of
interest so as to be neglected.
Since the values for the $r_t$ and GC center positions are 
adopted from the current version of the on line database of \cite{harris96}, the 
remaining free parameters for which we solve are the
central surface brightness ($A_0$) and the core radius ($r_c$) for each color.
The fitting procedure  to determine the surface brightness uses a
weighting scheme for each point (i.e. each value of $r_{eff}$)  
based on its uncertainty, and returns the value of each of the
two parameters and an error for each.  The Levenberg-Marquardt 
fitting algorithm was used as implemented in IDL taken from
http://cow.physics.wisc.edu/{\textasciitilde}craigm/idl/down/mpfit.pro.
All the
codes required to determine the surface brightness profiles
and to fit them were written by SH in IDL.

The minimum detection for a GC to be included here is a central surface
brightness of at least 10 DN
above the background with a clean detection in each of the first four annuli
in each of \irfilt. Of the 150 GCs in 
the current on-line version of \cite{harris96},
\ntotal\ are included in our sample.

\begin{figure}
\epsscale{0.9}
\figurenum{1}
\plotone{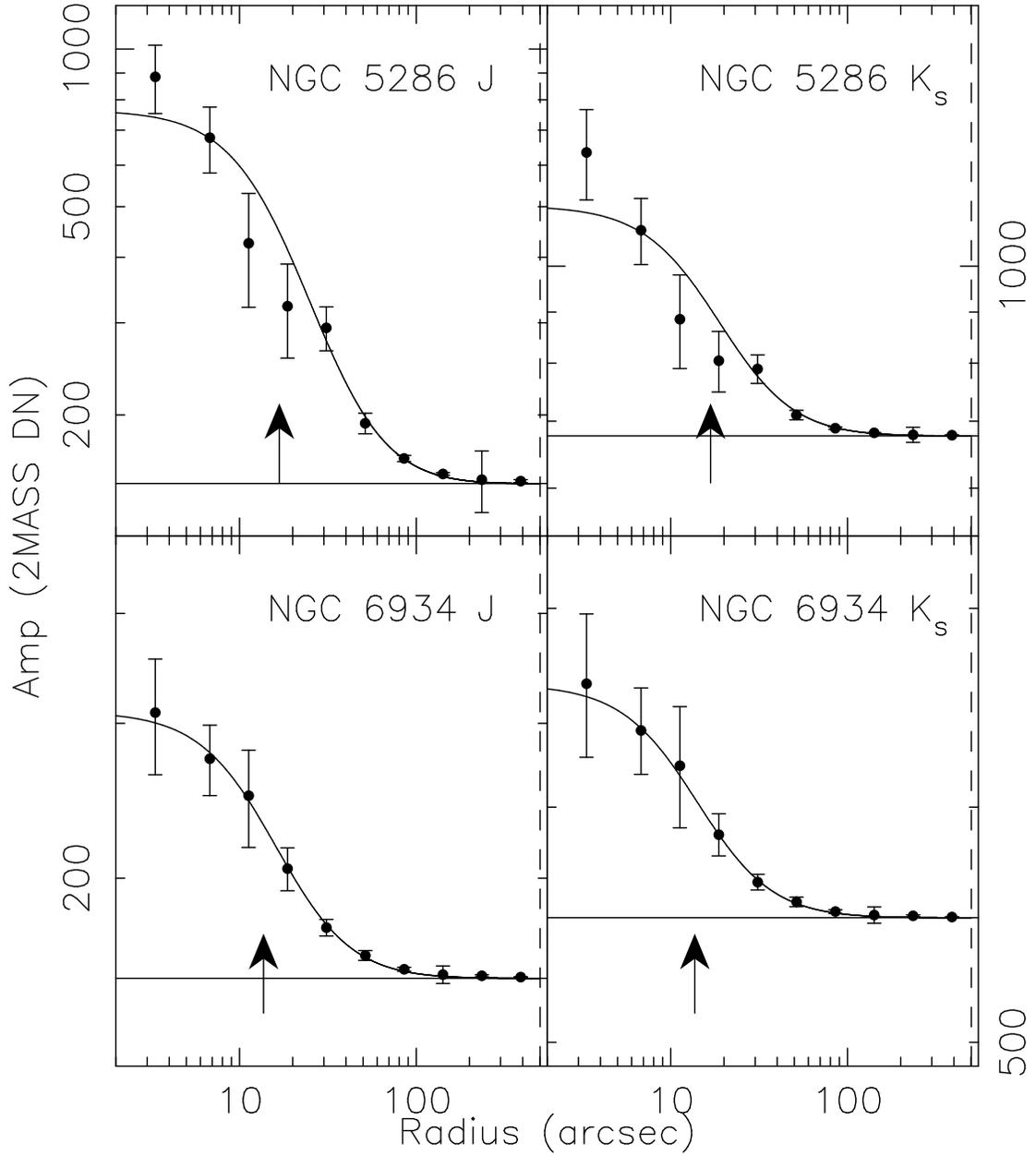}
\caption[]{The surface brightness for the 30th brightest (top) and 
30th faintest (bottom) 
GCs in our sample are shown for $J$ (left panels) and for $K_s$ (right panels).
The fit King profiles are also shown.  An arrow marks $r_c(J)$ and
a vertical dashed line indicates the tidal radius.  The horizontal line
indicates the background.
\label{figure_sb}}
\end{figure}

Fig.~\ref{figure_sb} shows our derived surface brightness profiles
for the 30th brightest and 30th faintest GCs in our sample at $J$ and at
$K_S$.  The fit
King profile is superposed.

\subsection{The Optical Surface Brightness Profiles}

We matched our IR surface brightness profiles derived from 2MASS images
onto optical ones to construct such colors as $V-K_s$.
The surface brightness profiles at $V$ were taken from the literature.
The primary source is TKD95. They have 
carried out an analysis of the extensive material collected
by the Berkeley Cluster Survey \citep[see, e.g.][]{george86}, as well as
compiled
many other sources of optical photometry,
particularly \cite{peterson86}. They  then 
fit the homogenized  set of data for each GC
with a grid of single mass, isotropic, non-rotating \cite{king66a} models.
Values of $r_c$, $r_t$
and central surface brightness
appropriate for the $V$ filter
were taken from their Table~2 when available.  The resulting
King profile was then
integrated out to the desired radius to obtain the integrated light 
at $V$ for a specified aperture.
If there was no data for a specific GC in TKD95, but the required parameters
were given in the on-line compilation database of
\cite{harris96}, the values there were used.
Nine of the GCs in our sample do not have a $V$ surface brightness
profile considered accurate from either of these two sources.

For all definite or possible core-collapsed
GCs as listed in Table~2 of TKD95,
we use the Chebyshev polynomial fits to the $V$ surface brightness
profile whose coefficients are given in Table~1 of 
TKD95.  Certain key information
about how to use these Chebyshev and polynomial fits
to the observed surface brightness as a function of radius 
is missing from TKD95 as published, as are the extensive and useful notes
to their Table~2.  These were kindly provided to us
by S.~Trager and are now available through a link on his home page, see
http://www.astro.rug.nl/{\textasciitilde}sctrager/globs/cheb\_transform.txt
and table2notes.txt.
Even with the aid of his notes, it was still difficult to make proper
use of the Chebyshev polynomial fits, and for \nvdirect\ of the GCs in our
sample, including all of the core-collapsed ones,  aperture photometry
was carried out by
integrating the measurements of the compiled $V$ surface
brightness profiles given in 
Table~1 of TKD95 (available through the on line edition of the AJ)
to derive integrated $V$ magnitudes.
A few GCs have surface  brightness profiles in this table from TKD95
with an arbitrary photometric zero point; these were ignored.

The optical surface brightness profiles for Galactic GCs
are dependent on a compilation of measurements 
from many different programs carried out by many different groups
utilizing different
telescopes, instruments, filter sets, standard star fields, etc.  
Their zero points are based on assuming
photometric sky conditions prevailed at some particular time.
They do not have the all-sky uniformity 
of the photometric zero points which 2MASS has.    
This critical difference means that the errors of any $V$
surface brightness profile
for the brighter GCs derived from reasonably deep digital (i.e. CCD)
image are dominated
by the uncertainty in its photometric zero point, not by random 
statistical measuring errors,
centering errors, or stochastic errors from the finite number of very bright
stars near the RGB tip.

\subsection{The Core Radii}

In our initial implementation, the core radii of the King
profiles were derived independently from the fit to each of the three 
filters \irfilt. 
Fig.~\ref{figure_rcore} shows the difference between
the $r_c$ deduced for various pairs of filters
divided by the uncertainty of this difference as a function
of the core radius of each GC at $V$.  A minimum uncertainty
of 1.0'' was assumed for
each of the $r_c$ (including that of $V$, whose 
uncertainty is not easily available from published material).
The largest differences (in units of $\sigma$) occur
among the core-collapsed
clusters (circled in the plots), where the optical $r_c$, 
determined from material
with better spatial resolution, are typically only a few arcsec,
and are always smaller than the
IR $r_c$ values.  There is
excellent agreement among the core radii determined from the various
IR filters.
The lower right panel illustrates this for the $J$ and $K_s$ filters;
the difference for each GC between $r_c(J)$ and $r_c(K_s)$,
normalized by the appropriate $\sigma$, is shown there.

The uncertainties in $r_c$ as listed in Table~\ref{table_rcore}
are typically a few arcsec, and adaption of the independently
determined $r_c$ for each of \irfilt\ led to unsatisfactory results,
including, for example, the presence of
many outliers in a plot of $J-K_s$ colors versus
[Fe/H].  The core collapsed GCs were among the worst
of the outliers, as might be expected given their small $r_c$.
In view of the excellent agreement among $r_c(J)$, $r_c(H)$ and
$r_c(K_s)$ for each cluster, we
decided to tie the IR values of $r_c$ together.
The set of  $r_c$ values determined from the $J$ mosaics of each GC
are presumably the most accurate among the three IR colors; 
the sky is much darker than at
$H$ or $K_s$, while the central surface brightness 
at $J$ of each GC is only slightly
smaller than in the other two IR filters.  In addition, the stochastic
effects are smaller at $J$ than in the redder filters.
We thus set the $H$ and $K_s$
core radii to the value obtained from the $J$ mosaic of each 
GC.  

\section{The Adopted [Fe/H] Values}

We adopt as our primary source of metallicities
the recent homogenized compilation 
by \cite{kraft03} of values for [Fe/H] 
for Galactic GCs based on detailed analyses of Fe~II lines from high dispersion
spectra of individual red giants.  In particular, we adopt the values
given in the last column of their Table~7, based on
Kurucz model atmospheres without overshoot 
\citep{kurucz93,castelli97}\footnote{Grids of Kurucz model atmospheres can be
downloaded from http://kurucz.harvard.edu/grids/html.}.
The deduced Solar Fe abundance for the work of
\cite{kraft03} is $\epsilon$(Fe) = 7.52 dex. They include
values based on observations of the IR Ca triplet
in individual GC red giants by \cite{rutledge97}, transformed 
onto their system, assuming
a linear transformation applies.  

There are still very few accurate determinations of metallicity for
the most metal rich Galactic bulge clusters.   We adopt
the results of \cite{cohen99} and \cite{carretta01}
for NGC~6553 and for NGC~6528, the archetypical
populous metal-rich bulge GCs with
the smallest (but still high) reddening values.  There are two
other GCs in our sample 
with [Fe/H] $> -0.2$ dex, Terzan~5 and Liller~1.  Their adopted high
metallicities are taken  from \cite{harris96}; neither is included 
in the compilation of \cite{kraft03}.  \cite{origlia02} and
\cite{origlia04}, who analyzed  high resolution near IR Keck spectra for luminous
giants in Liller~1 and in Terzan~5, confirm the very high metallicity
of both of these GCs.  The moderate resolution IR spectroscopy of individual
red giants in Liller~1 by \cite{stephens04} also supports a very high
metallicity for Liller~1.

All of the
GCs in the \cite{kraft03} compilation with metallicities higher
than that of 47 Tuc or M71 are in fact from \cite{rutledge97}.
The (high) Fe-metallicities we have adopted above for NGC~6528 and NGC~6553 
then suggest that the relationship
between IR Ca triplet line strength (the $W$' parameter of 
Rutledge \etal\ 1997) and [Fe/H] becomes non-linear at high [Fe/H], contrary
to the assumption made by \cite{kraft03}.  Of the 
high-metallicity GCs incorporated
into the compilation in this way, only NGC~6304
is probably affected (its [Fe/H] being underestimated)
at a level exceeding 0.1 dex.

Results from the extensive program of \cite{carretta97},
including high dispersion analyses of a sample of 24 Galactic GCs,
are not incorporated into the compilation
of \cite{kraft03}. \cite{carretta97} studied only one
GC with [Fe/H] $> -0.7$ dex not already included in the 
more extensive compilation of \cite{kraft03}, NGC~6352; their
derived [Fe/H] is within 0.05 dex of that from \cite{harris96}.
\cite{kraft03} do not include NGC~5272 (M3) in their compilation.
We adopt [Fe/H](FeII) from \cite{cohen05}, adjusted for the
difference in log$\epsilon$(Fe) for the Sun,  of $-1.36$ dex.

For the 64 GCs in our sample with no entry in the \cite{kraft03} compilation
or not specifically discussed  above, the [Fe/H] values given in the
current on-line database of \cite{harris96}, which 
are primarily from \cite{zinn84}, are adopted.  The  [Fe/H] values
we adopt and their sources are given for each GC in Table~\ref{table_colors50}.

\section{Forming the Colors}

We use the values of $E(B-V)$ values
given in H96 to remove the
interstellar extinction.
We adopt the reddening curve of \cite{cardelli89},
$A/E(B-V) = 3.10, 0.90, 0.58$ and 0.37 for
$V$, $J$, $H$, and $K_s$ respectively.  Integration of the King
profile fits out to a specified radius for each filter
$VJHK_s$ then produces the integrated light magnitudes 
of the Galactic GCs.  As discussed
above, due to problems in using the TKD95 polynomial fits, we 
directly integrated
the observed $V$ surface brightness measurements in many cases.
We also ended up doing this for the IR colors as well for most of the
GCs in our sample, as we will see below.

We divide the sample of Galactic GCs into three groups based on reddening
and on the accuracy of the central surface brightness at $K_s$ of the fit
King profile.  The ``best'' 
group has SNR($K_s) > 10$ and $E(B-V) < 0.4$ mag for $V-K$.
A larger
reddening can be tolerated for $J-K_s$ while still introducing
a fixed maximum uncertainty in the color
due to the much lower sensitivity of this color to a change
in $E(B-V)$.  Taking into account the size of the range
in color as well as the dependence of reddening on wavelength, 
we adopt a cutoff in $E(B-V)$ of 1.0 mag for
$J-K_s$.  The ``fair'' group has, for both of these colors,
the SNR limit reduced to 5.  The total sample studied here
is \ntotal\ Galactic GCs; the number in each group is given
in Table~\ref{table_n_samp}.

Since the values of $r_c$ and $r_t$ are fixed for \irfilt, the
photometric errors in IR-IR colors can be calculated directly from the
uncertainties in the central surface brightness (SB) found
from the King profile fits and the uncertainties in the background values.
This ignores other sources of errors such as an incorrect choice of
$E(B-V)$.  If we assume a 20\% uncertainty in $E(B-V)$,
then an uncertainty in $J-K_s$ of 0.1 mag results when 
$E(B-V) = 2.1$ mag; the reddenings of 5 of the GCs in our sample exceed
that value.  The centroiding error is
less important,
since the same center was adopted for each of \irfilt. 

This straightforward error calculation of the photometric errors leads to 
substantial uncertainties for 2MASS -- 2MASS 
colors for the fainter GCs. The $A_0$ values deduced from the King profile fit
are subject to centroiding
and stochastic problems, which increase the dispersion among the
8 sectors in each
radial annulus.  This increase in $\sigma$ increases the uncertainty
in the derived $A_0$, and hence the calculated uncertainty for the IR--IR colors.
Given the small range in IR--IR colors such as $J-K_s$, an uncertainty
larger than 0.1 mag is highly undesirable.  
For those GCs with ${\sigma}(J-K_s) > 0.15$ mag as calculated from the
King profile fit parameters, we therefore bypassed the King profile fits
and instead
directly integrated our own measured SB profiles from the 2MASS images
in \irfilt\ out to the desired radius of 50'' rather than integrating
the fit King profile.  This is equivalent to aperture photometry
with some censoring of the data to eliminate bright non-cluster members.
In the end, this was done for
for  essentially all (\nirdirect) of our sample of  Galactic GCs.
We used the means for the central
two annuli, and the medians for the outer annuli, so as to
eliminate bright field stars.  The uncertainties from 
statistical fluctuations in measurements in 
any 2MASS -- 2MASS color then become much smaller, as many pixels
contribute to each measurement.  Thus the SNR limits
of 10 and 5 adopted for the ``best'' and the ``fair'' samples
actually correspond to much larger values of SNR. 
Maximum errors in \irfilt\ using direct integration
become 0.15, 0.15 and 0.21 mag respectively.

Colors and their uncertainties for an aperture with a radius
of 50'' are given in Table~\ref{table_colors50}.
IR colors based on 2MASS images cannot be determined accurately
for apertures much larger  than the 50'' radius adopted here
due to the modest depth of these images.
All colors in this table are reddening corrected and on the 2MASS system.
There are no blue outliers; i.e. there are no GCs in our
sample with $(V-K_S)_0 < 1.7$ mag.
Only two of the GCs in our sample are red outliers, with 
$(V-K_s)_0 > 4.3$ mag; they lie beyond the
maximum $(V-K_s)_0$ displayed in all our figures.  They 
are Ton~2 and Djorg~1, both of which have $E(B-V) > 1.0$ mag
and neither of which can be regarded as 
well studied GCs with accurately determined
metallicities or reddenings.

The uncertainties in IR--IR colors derived from measurement of 2MASS images
that are obtained via direct integration of our SB profiles
are small as the expected random (assumed Gaussian)
fluctuations in measurement are small
and the all-sky calibration of 2MASS photometry eliminates 
errors in the photometric zero point.   However, many more terms
make substantial contributions to
the uncertainties in optical--IR colors such as $V-K_s$
in addition to the expected random (assumed Gaussian)
fluctuations in measurement.
Optical--2MASS colors are seriously affected 
by non-random errors specific to each GC
such as incorrect zero points for the $V$ photometry, 
incorrect choice of reddening, or perhaps
to a smaller extent inconsistent choice of the adopted cluster center
for the two filters forming the color.
We consider the contributions to the uncertainty in
an optical--IR color of two of these in detail.  

A check was made to determine how prevalent small errors
for the GC center locations taken from H96
might be.  We looked in the range of values of the background-subtracted
amplitudes for the 8 sectors in the inner two annuli of each GC, establishing
the ratio ($R$) of the maximum to the minimum value in each color and 
checking the position angle of the sector that gave the maximum value of $R$.
About 2/3 of those checked with adequate signal level (more than 100 DN
in each color in each sector of the central two annuli) showed systematic
evidence for a centroiding error, with $R > 1.4$ for at least 4 of the
6 possible combinations of filter and annulus, and
in addition showed agreement
in the sector position angle which gave the maximum signal between
the first and second annulus for
at least two of the three filters.  It is highly unlikely that 
contamination by field stars could produce this in regions
including and so close to the GC center.  Sampling (stochastic)
errors can also be ruled out as the culprit as even some of the brightest
GCs showed this. We therefore suspect
that small errors of a few arcsec in the published centroid 
location of Galactic GCs are common.  Examples of the implications of such
positional inconsistencies on the derived magnitudes and colors
are  shown in Table~\ref{table_center}.
The GCs included in that table are just those 
that early in the course of this analysis we happened to notice
might have centroiding problems.  We see that 
for two choices of cluster center location
separated by $\sim$5 arcsec integrating the fit King profiles
produces changes in optical--IR colors reaching 0.2 mag
(only up to 0.1 mag for IR--IR colors), while
integrating the IR SB directly produces  changes 
which are generally smaller for both the integrated IR magnitude
and the IR--IR colors.  This is another reason why we decided to use
direct integration in preference to integrating the fit King profile.

The second error source we consider in detail 
is potential stochastic effects due to the
small number of stars near the tip of the RGB, which dominate the light
in the IR.  This error term will be larger for optical--IR colors
than for IR--IR colors.  It is larger for more metal-rich GCs,
with their very cool and red stars near the RGB tip, than
for metal-poor GCs. We assume that photometric contamination
from non-members has been largely eliminated by the use
of the sector median for annuli with
$r_{eff} > 10$'', an assumption which may not be valid
for faint clusters at low galactic latitude, for which most
of the light may come from within 10''.  Here we evaluate
the  potential stochastic error in $V-K$ arising from
cluster members using the relationship for sampling errors of
\cite{king66b} and the luminosity function of M3.  The $V$ band 
luminosity function is
from \cite{sandage57}, while $V-K$ colors along the isochrone are taken from 
the grid of \cite{girardi02}.   At a total luminosity
of 1\% that of M3, the sampling error ($E$) in $V-K$  is 0.06 mag. 
The  fractional  sampling error for each GC is then
$E \propto \sqrt{L(M3)f(M3)\over{Lf}}$, where $f$ is the fraction of the 
total light
in the aperture of interest.  For the GCs studied here
and with our adopted aperture radius of 50'', the sampling error
in $V-K$
ranges up to 0.15 mag; the value for M3 itself is 0.01 mag.
This term will dominate the V--IR color error in a few cases, and will 
be comparable to the probable $V$ photometric zero point uncertainty
error in additional cases.  

Many other factors may also contribute to the optical--IR color
uncertainties for GCs. For those clusters with large
$E(B-V)$, the reddening  will surely be patchy over the face of the GC
\citep[see. e.g.][]{cohen95}.
Application of standard reddening corrections with any adopted
effective $E(B-V)$ cannot 
accurately reproduce the true reddening corrections
of such objects.  For the GCs with very high
background stellar density, such as those seen
against the Galactic bulge, the issue of field star contamination
may become important, although we have taken a number of steps to minimize
this.  It should also be noted that we have assumed circular isophotes,
a valid assumption for most GCs. \cite{white87} find that only 32\% of
Galactic GCs are flatter than $b/a < 0.9$ and 5\% are flatter than
0.8 (NGC~6273 being the flattest  of their sample of 100 GCs, with $b/a = 0.73$), 
so on the whole the Galactic GCs are quite round.

Bearing all this in mind, we ascribe to optical--IR colors
uncertainties of 0.20 mag for the ``best'' sample, and 0.25 mag for the
``fair'' sample of GCs considered here.  Even larger uncertainties
seem appropriate for the remaining GCs due to their high reddenings.

\begin{figure}
\epsscale{0.7}
\figurenum{3}
\plotone{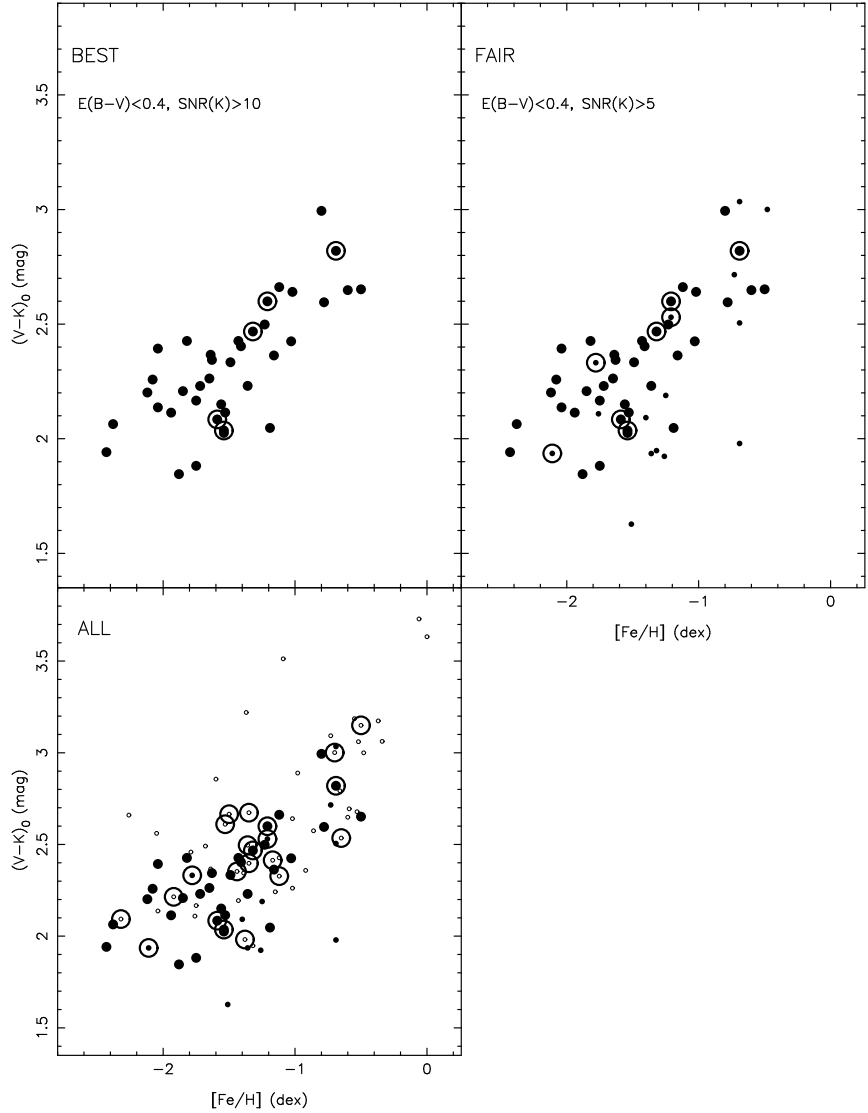}
\caption[]{Dereddened $V-K_s$  colors are shown as a function of 
[Fe/H] for the sample
of ``best'' (large filled circles), ``fair'' (the ``best'' sample
plus smaller filled circles) and ``all'' (adding in GCs denoted
by small open circles) GCs with IR 
surface brightness profiles from 2MASS
derived here.  Clusters which are, or may be, core collapsed (as indicated
in TKD95) are circled. An aperture 50'' in radius is used.
\label{figure_vmk_3panel}}
\end{figure}

\begin{figure}
\epsscale{0.7}
\figurenum{4}
\plotone{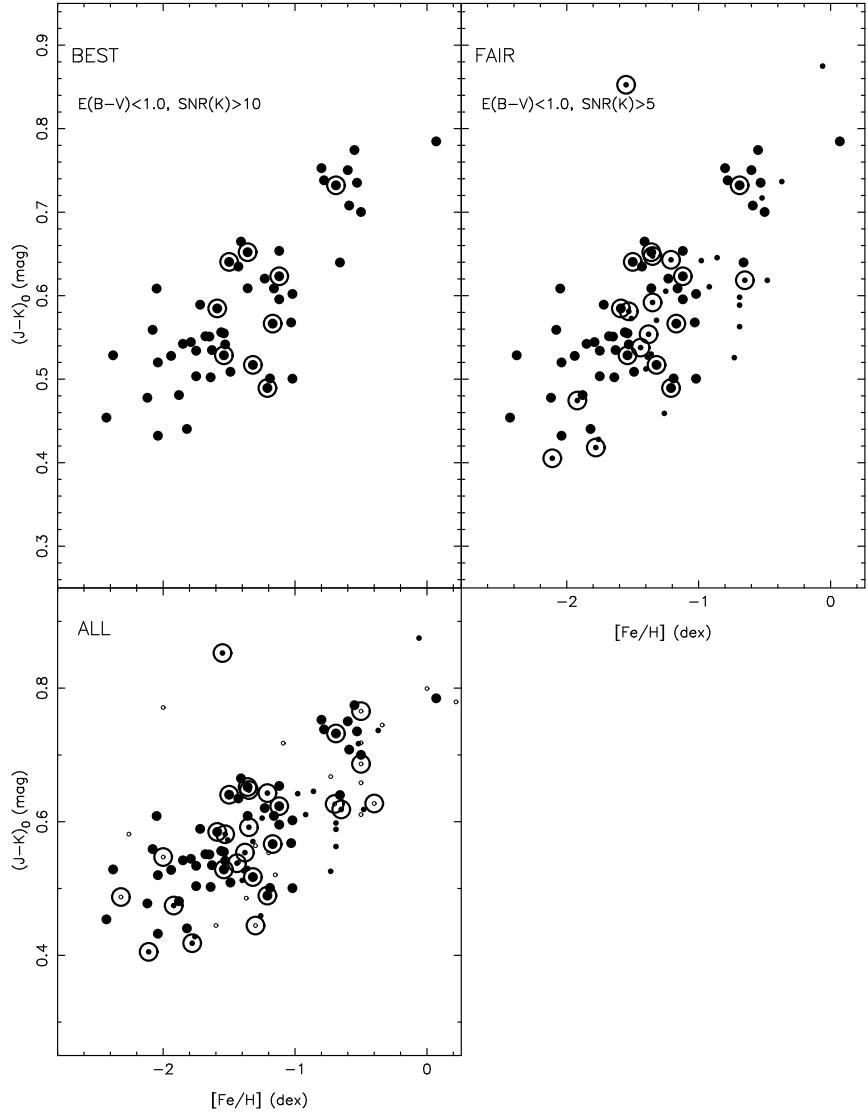}
\caption[]{Dereddened $J-K_s$ colors are 
shown as a function of [Fe/H] for the sample
of ``best'', ``fair'' and ``all'' GCs with IR 
surface brightness profiles from 2MASS
derived here.  The symbols are those of Fig.~\ref{figure_vmk_3panel}.  
An aperture 50'' in radius is used.  The red outlier appearing in the ``fair'' sample
is HP~1.
\label{figure_jmk_3panel}}
\end{figure}

Fig.~\ref{figure_vmk_3panel} and \ref{figure_jmk_3panel} show the 
reddening corrected
$(V-K_s)_0$ and
$(J-K_s)_0$ colors  as a function of [Fe/H] for the
sample of ``best'', ``fair'' and all included GCs.
The ``best'' sample is shown as large filled circles in the upper left panel.  
Smaller filled circles are used in the other two panels to denote 
those GCs in the ``fair'' sample not included in the ``best'' sample,
while  those not included in the ``fair'' sample 
are shown in the lower left panels as small open circles.
The $J-K_s$ plot shows a very good relationship, with small scatter.
Even the plot in the lower left panel including all \ntotal\ GCs in our
sample looks quite good.
The outlier in the ``fair'' sample is HP~1, which has a high uncertainty
in $J-K_s$ and lies
$\sim 2.5\sigma$ higher than typical for its
[Fe/H].  But in a sample this large, one such outlier might
be expected, and in addition this GC can hardly 
be considered a well studied cluster
with an accurately determined reddening or metallicity.

The $V-K_S$ plot (Fig.~\ref{figure_vmk_3panel}) 
has a much larger vertical scale than does Fig.~\ref{figure_jmk_3panel}.  
The relationship between $(V-K_s)_0$ and [Fe/H]
is good, but, not surprisingly, shows a significantly larger dispersion than that of
$(J-K_s)_0$ versus [Fe/H].  The lower left panel displaying the 96 GCs with
photometrically calibrated $V$ SB profiles has a very large $\sigma$ due 
in part to the
high $E(B-V)$ values of some of the GCs included here.
None of the four GCs in our sample with [Fe/H] $> -0.2$ dex is included in the
``best'' or ``fair'' $V-K_s$ sample; they each have reddenings
that exceed the cutoff value.

\section{Fits to V-J, V-H, V-K and J-K as Function of [Fe/H]}

We fit various $V-$IR and IR--IR 2MASS reddening corrected
colors as a function of
[Fe/H].  Quadratic fits are given for $V-J, V-H, V-K_s,
J-H$ and $J-K_s$.  Linear fits are given when there is little
improvement between the linear and second order fits.
A fit for the ``best'' and for the ``fair'' samples are carried
out for each color. The coefficients for these fits are
given in Table~\ref{table_feh_fit} as are the
rms dispersion about each fit of the sample GCs.  An  augmented
``fair'' sample was created, which additionally contains all of the four
GCs in our sample with [Fe/H] $> -0.2$ dex. 
This requires adding three GCs to the
$V-K_s$ ``fair'' sample (Liller~1 does not have a calibrated
$V$ surface brightness profile, hence no $V-K$ color).  Two GCs
(Terzan~5 and Liller~1) must be added to the sample for
$J-K_s$.  The uncertainty in $(V-K_s)_0$ is taken as
0.5 mag for Terzan~5, given its high reddening, and is set to
0.3 mag for the other two added GCs.  Uncertainties for the
four GCs with [Fe/H] $> -0.2$ dex in $(J-K_s)_0$ are assigned as 
a sum in quadrature of
the photometric error and the consequence of a 10\% uncertainty
in $E(B-V)$.    
Fits for the augmented sample in each color
(also given in Table~\ref{table_feh_fit})
enable us to probe the behavior of the
colors in the regime near Solar metallicity.

Linear fit are adequate for all $V-$IR colors unless
the very high [Fe/H] GCs are added, at which point quadratic
fits are clearly superior. Linear fits suffice for the
IR--IR colors of the GCs in the ``best'' sample.   But the quadratic
term is statistically  significant for the $J-K_s$ ``fair'' sample, which
already contains two of the four highest
metallicity GCs.

The dispersions around the fit
of the measured IR--IR 2MASS colors $J-K_s$ and $J-H$ are small
and only slightly larger than the expected assuming Gaussian
statistical variances of measurement for the observed signal levels.
Thus the many other potential sources of error are not of great
significance for these specific colors.  The rms dispersion
around the fits to the $V-$IR colors suggests typical total uncertainties
in the $V$ integrated light of $\sim$0.25 mag, in good agreement
with the estimates discussed above for the many 
terms contributing to the total error.

\begin{figure}
\epsscale{0.8}
\figurenum{5}
\plotone{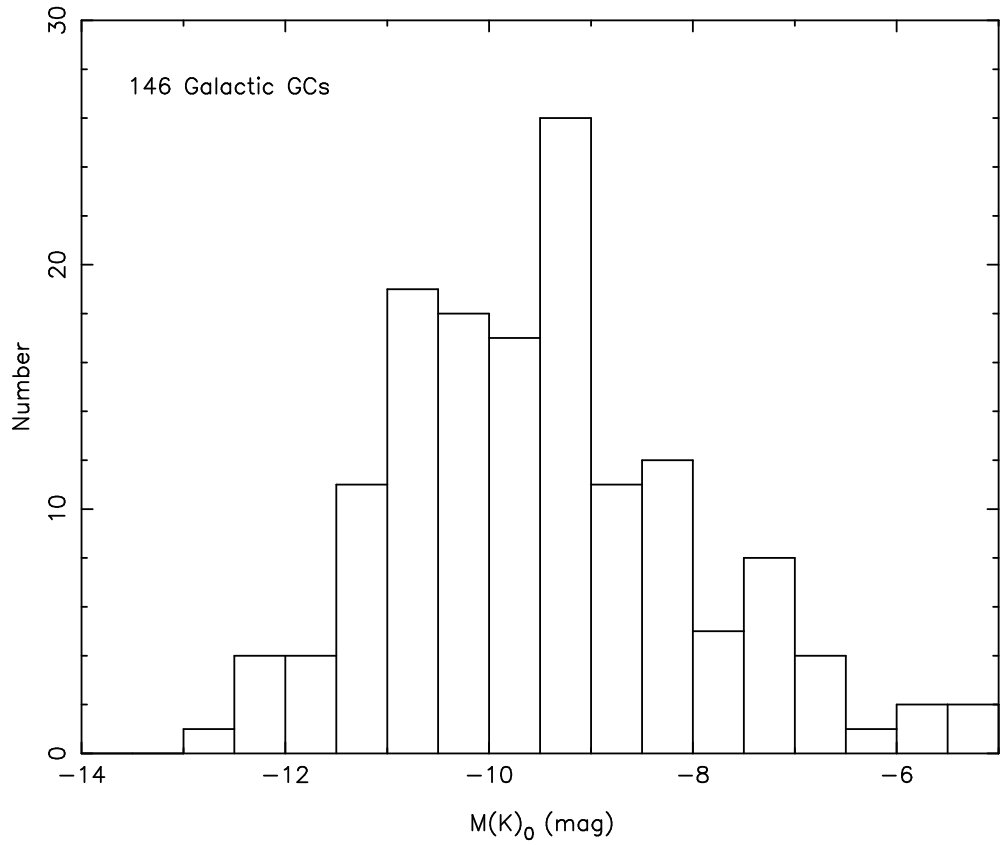}
\caption[]{The luminosity function at $K_s$ is shown for 146 of the Galactic 
globular clusters. \label{figure_lf}}
\end{figure}

The luminosity function for the Galactic globular cluster system 
at $K_s$ has been formed
by combining our $(V-K_s)_0$ colors with the total absolute $V$ mags
from the database of H96 for those GCs in our sample with $E(B-V) < 0.4$ mag.
For the remaining GCs,
the fits to $V-K_s$ as a function of [Fe/H] given
in Table~\ref{table_feh_fit} have
been used to predict the integrated light color from the Fe-metallicity
of each GC (taken from H96).  All 146 GCs from the
H96 database which have total $M_v$ tabulated there
are included.  There are perhaps another 5 known
Galactic GCs, all of which are extremely reddened and
poorly studied.  Fig.~\ref{figure_lf} shows the resulting $K_s$
luminosity function, which is 
peaked at $M(K)_0 \sim -9.7$ mag, or 
$L \sim 1.6 \times 10^5 L$\subsun\  for 
$M_K$(2MASS) $= 3.29$ mag, and (adopting $M/L_K = 1.4$)
is $M \sim 2.2 \times 10^5~M$\subsun.

\section{Comparison with Other Studies}

\subsection{The 1977 Data of Aaronson, Malkan \& Kleinmann
\label{section_acmm}}

The only previous substantial body of photometry of 
the integrated light of Galactic GCs in the IR
is the work of Aaronson, Malkan \& Kleinmann in the late
1970s.  A brief description of their data is given in 
\cite{acmm}, where the data was used in a number of plots.  However,
due to M. Aaronson's tragic and untimely death, the data
was never published in full\footnote{The data from M.~Aaronson and
M.~Malkan tabulated in \cite{brodie90} were unofficial preliminary
values for a subset of the clusters included in the 1978 study.}.  
%
%
We do so here in 
Table~\ref{table_acmm}, recognizing again that this is the data
of Aaronson, Malkan \& Kleinmann as it existed in 1978.  
They observed the central regions of 54 GCs
using a single channel photometer on the KPNO No. 1 0.9m telescope
with a beam size whose diameter in most cases was 105''.  Background
corrections were made chopping to fields $\sim$200'' away.  Integration
times were set to achieve a photometric accuracy of $\leq0.02$ mag
for $J$, $H$ and $K$.  Narrow band indices measuring the 
absorption in the 2.4$\mu$ CO band
and in the 1.9$\mu$ H$_2$O band were obtained as well for some
of these GCs.  They combined these with optical
surface brightness profiles from the literature
as it existed at that time to derive $V-K$ colors as well.
It is important to note that they used a smaller telescope than did
2MASS, with a now obsolete and noisy single channel detector, but with longer
integration times.
They divided their final sample into 14 calibrating GCs, whose reddenings
and metallicities were believed to be well known, 23 other GCs,
which were believed to be useful, and 27 GCs with only
one or two measurements, regarded as less reliable, which were not
used in \cite{acmm}.

Two versions of this old data exist.  The first is a list of the observed colors,
preserved by M. Malkan from about 1977 and recovered from old computer
files.  These are the values given in Table~\ref{table_acmm}.
The observed broad band colors are listed, while the
reddening corrected CO and H$_2$O indices are tabulated.  The
reddening corrections for the narrow band indices are very small
as the wavelength range covered in these measurements is very narrow.
\cite{frogel79} use $E(CO)/A_V =-0.007$\footnote{The reddening corrected CO indices
are larger than the observed ones.} and $E(H_2O)/A_V = 0.019$ mag, so
any difference between the $E(B-V)$ values adopted in 1978
versus those in current use has a negligible effect.
The second archive of these integrated light GC observations
is a list preserved in a notebook from 1977
of the dereddened values used by J. Cohen to generate the figures and fits
presented in \cite{acmm}.   These
values agree well with those in M. Malkan's
archive for $J-K$ (the mean difference for 37 GCs is 0.01 mag,
with $\sigma = 0.04$ mag) with somewhat larger differences in $V-K$ 
the (mean difference for 35 GCs is 0.04 mag, with
$\sigma = 0.11$ mag).  It is believed that these differences
arise from the slightly different values of $E(B-V)$ 
and in the mean colors used in 1977 during the preparation of the
manuscript for
\cite{acmm} versus those adopted and archived by M. Malkan at 
the end of all relevant observing runs and reduction thereof in 1979.
The four GCs with $\Delta(V-K)$ exceeding 0.20 mag 
between the two independent archives are marked in
the table.  The nominal errors of these measurements, excluding
the 27 considered less reliable, henceforth ignored here,
are  $\pm$0.15 mag for $V-K$ and $\pm$0.04 mag for $J-K$.

In order to compare our colors derived from 2MASS and those
of Aaronson, Malkan \& Kleinmann as they existed in 1978, 
we transform
the observed colors recorded by M. Malkan 
from the CIT system to which we believe
the measurements were calibrated\footnote{We thank
the referee,  John Huchra, and
Jay Frogel for confirming that the CIT system was used.} into that 
of 2MASS using the equations in \S4.3 of \cite{carpenter01}.

\begin{figure}
\epsscale{0.8}
\figurenum{6}
\plotone{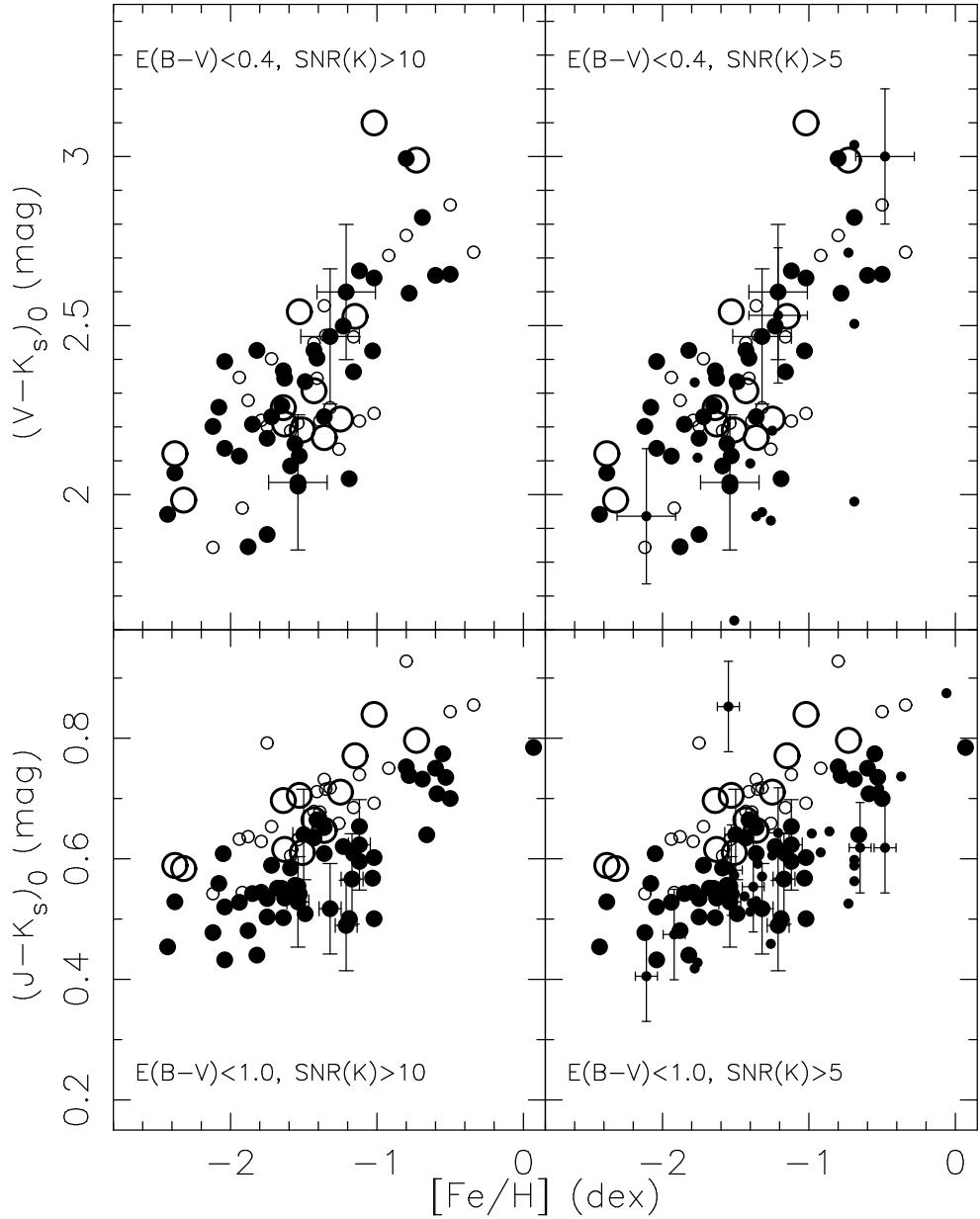}
\caption[]{The top panels show $(V-K_s)_0$ versus [Fe/H] from our
``best'' (left, large filled circles) and ``fair'' (right, small
filled circles) samples; the bottom panels show
the same for $(J-K_s)_0$. A 50'' radius circular aperture is used. 
Core collapsed GCs are marked 
with crosses.  The 1978
data of Aaronson, Malkan \& Kleinmann, transformed as
described in \S\ref{section_acmm}, is superposed (large open circles
are their calibrating clusters
which they believed to have accurate metallicities and reddenings; 
smaller open circles are other GCs used in \cite{acmm}.
\label{figure_4panel_acmm}}
\end{figure}

\begin{figure}
\epsscale{0.9}
\figurenum{7}
\plotone{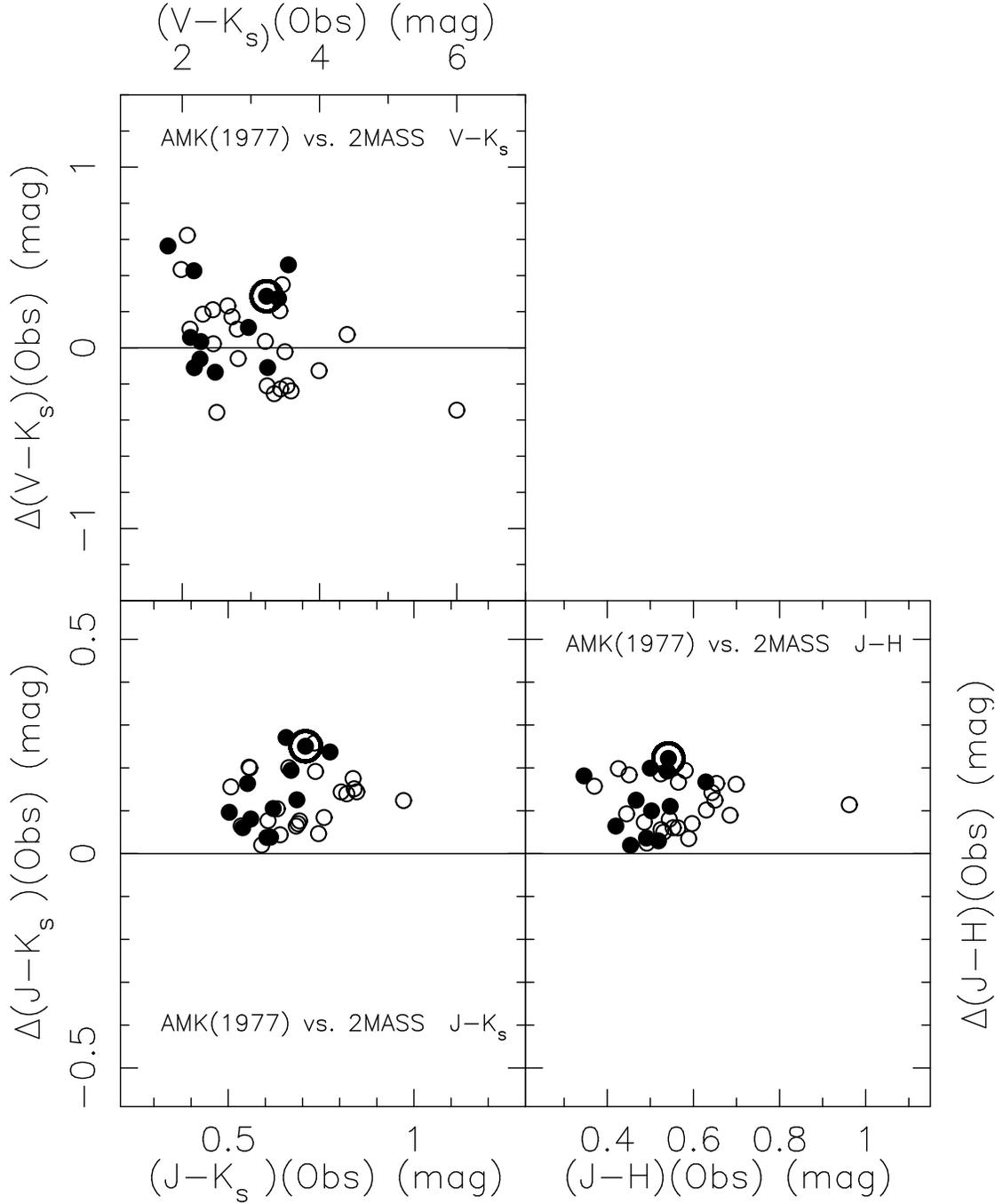}
\caption[]{The difference between the 1978 colors of 
Aaronson, Malkan \& Kleinmann and the 2MASS colors presented here,
as observed and in the 2MASS system for both, are shown for
$V-K_s$,  $J-K_s$ and $J-H$ as a function of our 2MASS colors.
\label{figure_acmm_comp}}
\end{figure}

We show in Fig.~\ref{figure_4panel_acmm} 
our ``best'' and ``fair'' samples
in the reddening corrected colors  $(V-K_s)_0$ and in $(J-K_s)_0$ as a function of
cluster Fe-metallicity with the results of 
Aaronson, Malkan \& Kleinmann superposed.  Current values
for the reddening and metallicity for each GC are used
with the 1977 observed colors in this figure. 
The range
of GC colors is much smaller in $(J-K_s)_0$ than it is in $(V-K_s)_0$;
the scale of the Y axis of
the upper panel of Fig.~\ref{figure_4panel_acmm}
is correspondingly much larger than that of the lower panels. 
The differences are shown as functions
of our derived 2MASS colors in
Fig.~\ref{figure_acmm_comp}; some statistics of these differences
are given in Table~\ref{table_acmm_comp}. Observed colors are compared here;
the choice of $E(B-V$ and of metallicity for each
GC is irrelevant. This table
shows that the dispersion in the differences for $V-K_s$
as measured in 1977 (transformed into the 2MASS system)
and our measurements is consistent with the errors, and the means
agree to within the uncertainties of the measurements.
For $J-H$ and $J-K_s$, the dispersion in the differences between the
colors of
Aaronson, Malkan \& Kleinmann 1977 photometry (transformed
into the 2MASS system) and our colors is small, only 0.07 mag,
easily consistent with the measurement uncertainties for the
two data sets. 
However, there is a small systematic offset, apparent in both
Table~\ref{table_acmm_comp} and in the lower panels of
Fig.~\ref{figure_acmm_comp},
of 0.13 mag such that the 1977 colors are systematically
redder in $J-K_s$ and in $J-H$ than our colors.  This does not appear to be
function of $J-K_s$, but rather a constant offset.

We ascribe these  systematic offsets in the IR--IR colors
between the 1977 data and the present set, at least in part,
to the difficulty of tracing now exactly how
the 1978 measurements were calibrated and
of transforming between the various photometric systems involved.  
The $J$ filter adopted by the 2MASS 
project is somewhat broader than most other $J$ filters, extending
into the adjacent blue and red H$_2$O absorption bands; see the
discussion in \cite{carpenter01}, who has derived
relationships between the many flavors of $JHK$ in use  and the filter
set adopted by 2MASS, and in \cite{2mass2}.  When one examines the
range of the coefficients
for transforming various types of $J-K$ colors into the 2MASS system
over the full suite of IR photometric
systems in use, one concludes that it might be possible to explain 
the small systematic offsets seen in the lower panels
Fig.~\ref{figure_acmm_comp}
and in Table~\ref{table_acmm_comp} for 
$J-K_s$ and for $J-H$  as  errors in the
coefficients of the transformation equation we used.
The definition of the $H$ and of the $K$ filters are
more consistent between the various IR photometric systems in use
than that of $J$  and hence $H$ or $K$ magnitudes
are less subject to such transformation uncertainties.

The coefficients of the fits to $V-K$ and $J-K$ versus [Fe/H]
derived by \cite{acmm} in 1977  (transformed into
2MASS colors)
from their small sample of
calibrating GCs (only 14 clusters) are
included in Table~\ref{table_feh_fit}.
A comparison of these linear fits with those to the ``best'' present
measurements versus [Fe/H]
shows excellent agreement in both cases, as should be expected
given the agreement between the two data sets shown in
Fig.~\ref{figure_4panel_acmm}.
The constant coefficients for $J-K_s$
differ by only 0.03 mag, well within the errors of the 
1977 fit, with the 1977
data being slightly redder for a fixed [Fe/H], as 
expected from the discussion above.

Given the uncertainties of the Aaronson, Malkan \& Kleinmann
data and of the present colors derived from 2MASS images,
the agreement overall is very good for $V-$IR(2MASS) and reasonably good
for IR--IR colors. 
We have  demonstrated that the measurements of 
integrated light colors of Galactic GCs carried out by
Aaronson, Malkan \&
Kleinmann in 1977 appear to be valid and to agree 
reasonably well with our current
measurements based on 2MASS images.  This suggests that the 
overlap found by \cite{frogel80} between the integrated light
colors of the M31 GCs and those of the Milky Way GCs is
also still valid.  This might not hold for the expanded set of objects
today considered to be GCs in M31, but see the discussion 
regarding the reliability of identifications of
purported ``young GCs'' in M31 by \cite{cohen05b}.

\subsection{The Work of \cite{nantais06} }

\begin{figure}
\epsscale{0.9}
\figurenum{8}
\plotone{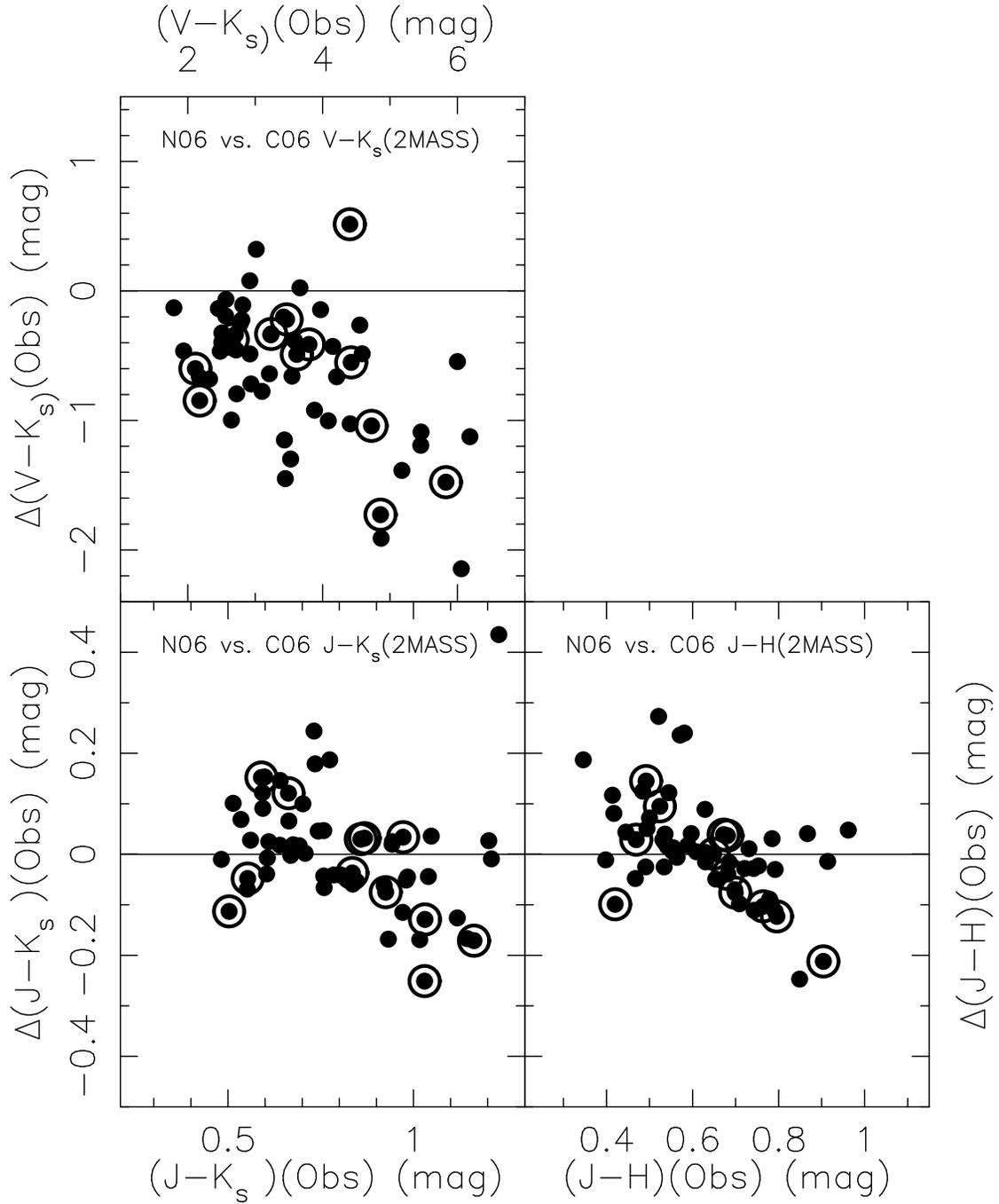}
\caption[]{The difference between the colors of
\cite{nantais06} and those presented here,
as observed and in the 2MASS system for both, are shown for
$V-K_s$ (upper panels) and for $J-K_s$ (lower panels)
as a function of our observed 2MASS colors.  Core collapse GCs
are circled.
\label{figure_huchra_comp}}
\end{figure}

Very recently \cite{nantais06} present a compilation of
infrared light photometry for 96 Galactic GCs generated using aperture photometry
for diameters from 7 to 70 arcsec on 2MASS images.
This was combined with integration of the optical surface brightness measurements
of \cite{peterson86} to create $V-K_s$ colors.  Of these
only 68 were considered reliable, the remainder having
problems in the matching of the optical and IR photometry.
A comparison of our results with theirs is 
shown in Fig.~\ref{figure_huchra_comp} 
for the sample in common, with statistics of the differences 
given in Table~\ref{table_huchra_comp}. Those with
$V-J < 0$, considered not ``reliable'' by \cite{nantais06},
were excluded.  They  also exclude those with $V-J < V-I$;
J.~Nantais kindly supplied a list of those GCs excluded
as not ``reliable''.

Figure~\ref{figure_huchra_comp} illustrates the differences
in the two sets of colors for the integrated light of Galactic GCs.
The agreement in the mean for these two data sets
for the pure IR colors, i.e. $J-K_s$ or $J-H$,
is excellent and  the  dispersion of the set of differences (0.12 mag)
is consistent with the
photometric errors we have calculated
(given in Table~\ref{table_colors50}).  Since both sets are based
on 2MASS images, this agreement, while gratifying, is only to
be expected.
A comparison of the two sets of $V-K_s$ colors for the objects
in common, however,
shows a very large mean difference of 0.63 mag (ours being on average
redder) and a
very large dispersion ($\sigma = 0.52$ mags).  This is quite unlike
the comparison of our $V-K_s$ colors for Galactic GCs with
those of Aaronson, Malkan \& Kleinmann from 1977;
compare the upper left panel of Figure~\ref{figure_huchra_comp}
with that of Figure~\ref{figure_acmm_comp} and note the much larger range
of $V-K_s$ shown in the latter figure.
Our $V-$IR colors
have been derived with some care and are in good agreement
with those of Aaronson, Malkan \& Kleinmann from 1977.   We
believe that they are more reliable than those of \cite{nantais06}.

\section{Color Gradients Within Globular Clusters}

Color gradients within 1$r_c$ of the center of a GC can
arise due to stochastic effects of the small number
of luminous giants dominating the IR light.  If these
are by chance not symmetrically distributed about the
true center of the total integrated light of the GC,
a distortion of the central position will occur.  
This will result in a small central region which is apparently redder 
and more luminous than expected based on the cluster surface
brightness profile over a large radial range.
For sparse clusters, there may be a statistical fluctuation
in the distribution of the most luminous red giants such that
there is no such star close to the location of the optical
center; a center bluer than the
integrated cluster light would then occur.  However,
we are interested here in possible
larger scale intrinsic gradients of the cluster light.
While our data are not ideal for this purpose given the
short exposures and relatively shallow depth of the 2MASS images,
we explore this issue.

\begin{figure}
\epsscale{0.9}
\figurenum{2}
\plotone{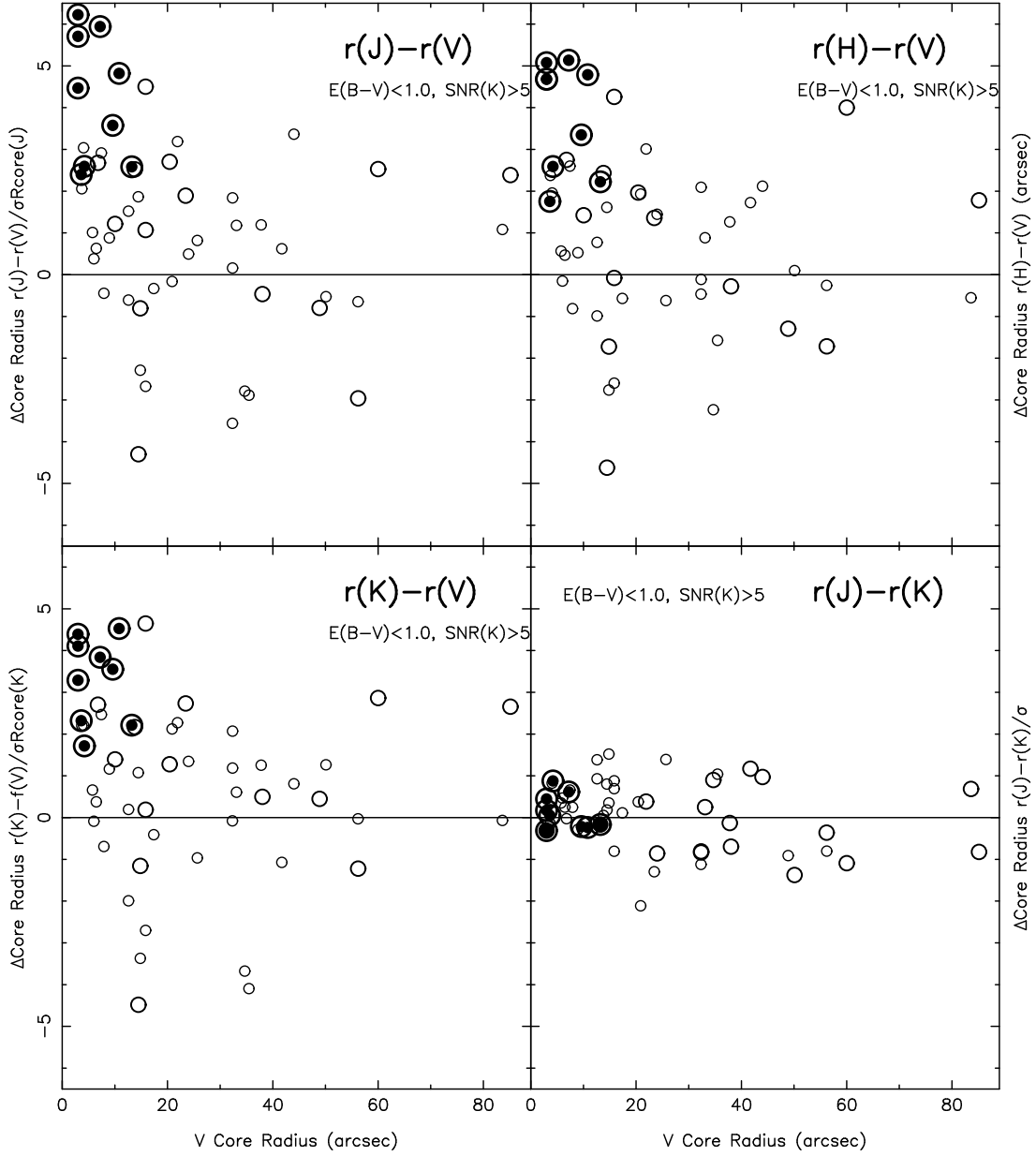}
\caption[]{The difference in core radius between $J$, $H$ or $K_s$ and $V$
divided by the uncertainty of this difference
is shown as a function of the $V$ core radius.  The lower right
panel shows the case of $r_c(J)$ as compared to 
$r_c(K_s)$.  A minimum uncertainty in each core radius of
1.0'' is assumed.  Probable core collapsed GCs are circled; their $r_c(V)$ are
from H96; all others are from TKD95.
\label{figure_rcore}}
\end{figure}

Our analysis suggests that $r_c$ is the same for each
of \irfilt\ to within the errors, as is demonstrated for $J$ and $K_s$ in 
the bottom right panel of Fig.~\ref{figure_rcore}.
We have explicitly assumed
in the construction of the surface brightness profiles for
\irfilt\ from 2MASS frames
that $r_t$ and $r_c$ are fixed for each GC.  This in
turn implies that IR--IR colors for Galactic GCs are independent
of radius, depending only on the ratio of the central
surface brightness in the two colors.  Furthermore, we consider
the IR surface brightness profiles as uncertain
at radii approaching $r_t$.  Thus only the
optical--IR colors among those considered here
could potentially reveal color gradients, and those only over
a radial range extending out to $r << r_t$.
Given our assumptions, the existence of a color
gradient in $V-$\irfilt\ would manifest itself in our analysis as
a difference 
between $r_c$ for $V$ and that for the 2MASS filters.
If $r_c(V)$ is larger than $r_c(J)$, the value to which
we set $r_c(H)$ and $r_c(K_s)$, then the integrated $V-J$
will become bluer as $r$ increases over
the radius range from 0 out to
about 5$r_c$, after which the color gradient is
not easily detected, since $r_t$ was assumed to be the
same for all colors investigated here. If $r_c(V)$ is smaller than $r_c(J)$,
$V-J$ will become redder as $r$ increases over that radial range.

The existence of color gradients, at the level to which we can
detect them, thus depends on whether there
are GCs for which $r_c$ is not the same for $V$ and for $J$.
Fig.~\ref{figure_rcore} shows that for most GCs, the assumption
of equality is valid.  This figure was constructed assuming that
the error in $r_c(V)$, which we do not know, is 1.0''.  If we
raise that to 2.0'', then only 11 GCs may show a detectable color
gradient.  Several of these are probable or definite
core collapsed GCs as indicated below.
NGC~1904 (c?) 4833, 6266 (c?), 6397 (C), 6522 (C) and 6356 have
$r_c(V) - r_c(J) > 2.5\sigma(\Delta{r_c})$, 
while NGC~6333, 6584 and 7006, with the same 2.5$\sigma$ tolerance
have $r_c(J) > r_c(V)$.  Only NGC~6266 (c?) and 6397 (C) have
a difference exceeding 4$\sigma$; these are a definite core-collapsed
GC and a probable one, so large differences in $r_c(V)$ versus
$r_c$(IR) should be expected given the spatial resolution of the 
2MASS images.  

The accuracy of the set of values
of $r_c(V)$, which
we have assumed here to be high (i.e. $\leq$2''), is crucial to this argument.
Yet the very recent work of \cite{ngc6266}, who determined an accurate
$V$ surface brightness profile for the
cluster NGC~6266, demonstrates that concern 
with the accuracy of values of $r_c(V)$ in  compilations
such as H96 is warranted.  Their recent
precision measurement of $r_c(V)$ for this GC is 19'', 
in agreement with our
value of $r_c(J)$ of $25.6\pm2.9$'', but is 30\% larger than
the value given by H96.   This resolves one of the two
cases for which a discrepancy of 4$\sigma$ or larger 
appears to exist between
the $V$ and IR core radius.   Furthermore, although
TKD95 called NGC~6266 a probable core collapsed cluster, \cite{ngc6266} 
find that NGC~6266 is not a core-collapsed GC.

Thus, to the level at which we
can detect color gradients and out to a radius $r < 5r_c$,
no GC in our sample appears
to have such,
but our ability to detect radial color gradients is 
severely limited by the modest depth of the 2MASS images.
Intrinsic large scale color gradients in GCs such as might
arise from mass segregation are difficult to detect even
in the best available data
as discussed by \cite{george93}.

\section{Comparison with Single Burst Integrated Light Models}

We next compare our results to a number of predictions from single
burst simple stellar populations (SSPs) of a unique age and metallicity.
There are many predicted grids of colors for SSP populations
based on various stellar evolutionary codes, assumptions about the
HB, the AGB, etc.  We must know which photometric system
was used to generate the model output colors as well as
that of any  photometric
databases used to calibrate the model's photometric zero points.

The specific SSP models considered here are those of
\cite{buzzoni89}, \cite{maraston05} and \cite{worthey94}.
A  somewhat larger consensus value for the age of GCs
was prevalent in the astronomical community prior to 2000, 
so we adopted models with ages of 12 Gyr from  
from \cite{buzzoni89}, while the 11 Gyr model of \cite{maraston05}
was selected.  We follow the assumption made by \cite{worthey94}
of an age of 15 Gyr for Galactic GCs.  These all use
the IR filter transmission curves of 
the Johnson system.  We
convert their predicted colors to the 2MASS system using
the transformations given in Appendix A of \cite{carpenter01}.

The [Fe/H] values we adopt here refer to Fe itself.
No adjustment has been made for any enhancement of the $\alpha$-process
elements, ubiquitous among GC stars.  Since it is generally believed
that [$\alpha$/Fe] $\sim +0.3$ dex for GC stars 
\citep[see, e.g.][and references therein]{cohen05}, we use the global metallicity
parameter as defined by \cite{salaris93} to adjust [Fe/H] to [M/H] 
(i.e. the parameter log(Z) used by stellar evolution codes).  Then
[M/H] = [Fe/H] + 0.2 dex
for the $\alpha$-enhancement typical of  GCs.
Each of the three model tracks were offset by $-0.2$ dex in
[Fe/H] to compensate for their assumed scaled Solar elemental abundances.

\begin{figure}
\epsscale{0.9}
\figurenum{9}
\plotone{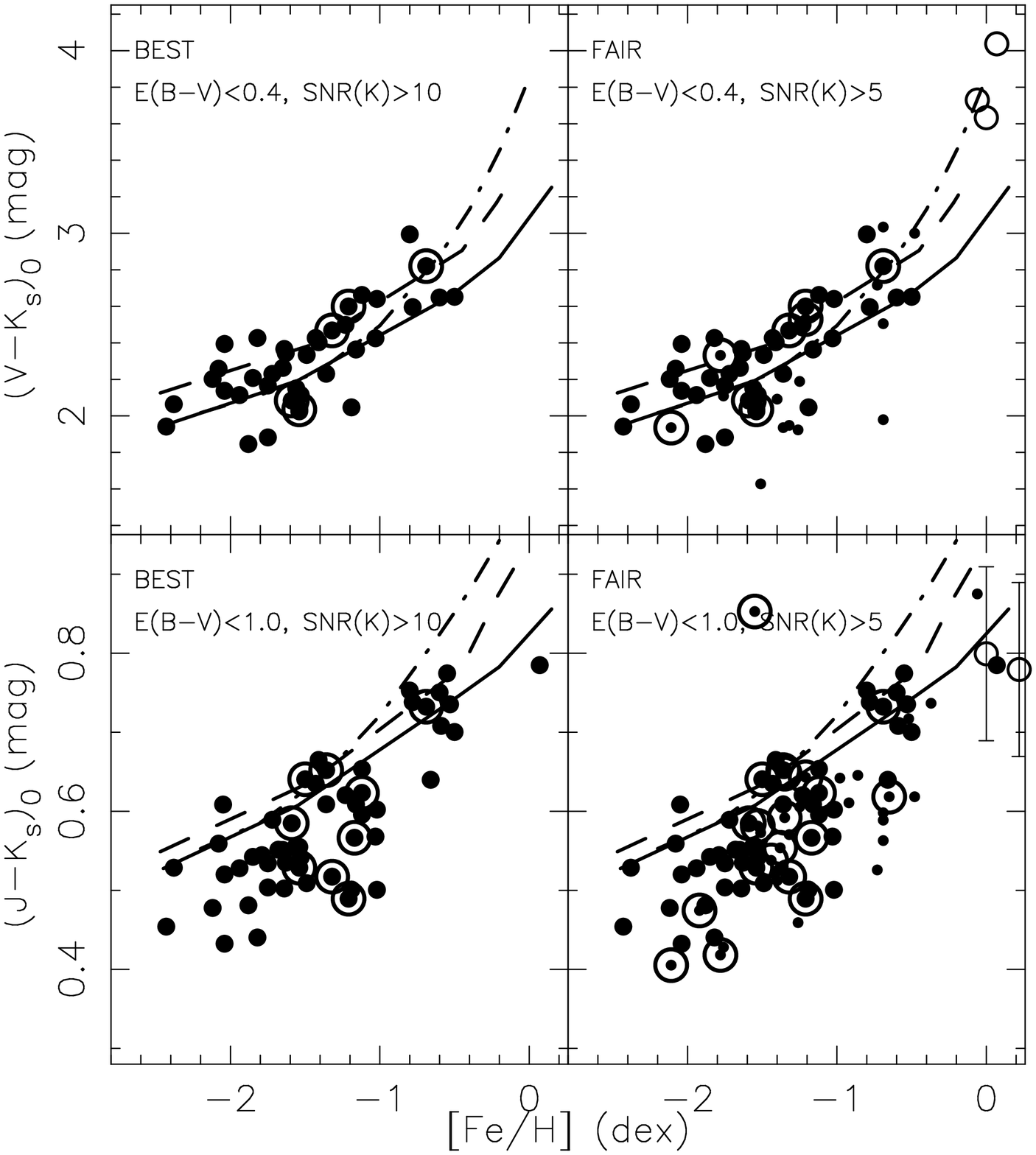}
\caption[]{Our derived dereddened
2MASS integrated light colors $(V-K_s)_0$ and $(J-K_s)_0$
are shown
as a function of [Fe/H] for the ``best'' and ``fair'' Galactic GC samples.  
Predicted SSP colors of a 11 Gyr model from 
\cite{maraston05} (solid curve) and a 15 Gyr model from
\cite{worthey94} (dot-dashed curve), as well as a 12.5 Gyr model of
\cite{buzzoni89} (long dashed curve) are superposed.  These have been
transformed from the Johnson into the 2MASS system and [Fe/H] values 
for each of the model curves have been adjusted
for the $\alpha$-element enhancement characteristic of Galactic GCs.
Additional GCs with [Fe/H] $> -0.2$ dex
which do not meet the criteria for the ``fair'' sample are shown in the
right panels as open circles. Error bars are shown for the two of these
which are heavily reddened in the lower right panel, see text for details.
\label{figure_model_comp}}
\end{figure}

Fig.~\ref{figure_model_comp} shows the three predicted SSP
model tracks superposed on the
``best'' and ``fair'' GC samples; the photometric system of
the displayed colors is that of 2MASS.  
Although for $V-K_s$ none of the four GCs
in our sample with [Fe/H] $> -0.2$ dex is in the ``fair'' sample,
the three of these clusters which have $V-K_s$ colors
are shown as open circles in the upper right panel.
The two such GCs (Terzan~5 and Liller~1) that are not in the $J-K_s$ ``fair''
sample due to their very high reddenings are similarly shown in the lower
right panel with error bars representing the sum in quadrature
of the photometric error and a the consequence of a 10\% uncertainty
in $E(B-V)$.

Each of the predicted SSP
color-metallicity tracks overlays the $(V-K_s)_0 -$ [Fe/H] relationship
we have derived for Galactic GCs over the metal-poor regime
[Fe/H] $< -0.5$ dex. 
However, the lower panels of Fig.~\ref{figure_model_comp} show
a slight offset of $\sim$0.1 mag at a fixed [Fe/H] such that the
models are slightly redder in $J-K_s$
than the color we derive from 2MASS.  Since the agreement of our
derived $J-K_s$ colors with those of \cite{nantais06} is perfect
(at the level of $\pm0.01$ mag), one cannot ascribe this difference
to problems in our IR--IR colors.

There are two possible explanations for this offset between the
SSP models and the actual colors of the Galactic GCs.
Fig.~\ref{figure_acmm_comp} and Table~\ref{table_acmm_comp}
show a similar small offset
between the photometry of integrated light colors by
Aaronson, Malkan \& Kleinmann from 1977 and the 2MASS based colors
presented here 
in the sense that
the 1977 $(J-K_s)_0$ colors transformed to the 2MASS system
are somewhat redder than the ones we derive here.  This is the
same sign as the differences seen between the predicted SSP model colors
and those we derive for Galactic GCs, and is of the same magnitude
as the problem seen in the lower panels Fig.~\ref{figure_model_comp}
for $J-K_s$. This should not be surprising, as
the validity of such models for the integrated light
of simple stellar systems as a function of metallicity
and age is generally established at least in part by attempting to reproduce
as a key test
the integrated light colors of Galactic GCs;  the ACMM colors
(i.e. their fits of color as a function of [Fe/H])
were the only ones available for this purpose prior to the present.
(This does not explain the origin of the offset in $J-K_s$ seen in the 
lower panels of Fig.~\ref{figure_model_comp}; it just shifts the problem 
back to the details
of the calibration  of the Aaronson, Malkan \& Kleinmann 1977 data.)

A second possibility relates to the photometric system of the
calibration data.
These models are all calibrated using the photometry of
\cite{frogel83} and \cite{cohen83} for individual red giants
in Galactic GCs, as that was the largest
sample of such data available until quite recently. 
Most model codes utilize the Johnson
filter transmission curves for $J$ and for $K$, while the key 1983 
data sets were calibrated to and published in the CIT system.  
Consider a star with a $(J-K)_0$(2MASS) color of 0.65 mag.  As observed,
for a typical $E(B-V)$ of 0.4 mag, it will have a color
on the 2MASS system of 0.86 mag, and will have an observed
$(J-K)$(CIT) of 0.83 mag while
$(J-K)$(Johnson) will be 0.90 mag
according to the transformation equations derived by
\cite{carpenter01}.  This star will thus be 0.07 mag redder
in $J-K$ in the Johnson
system than in the CIT system.  If a model code does not take these
differences among the IR systems into account, 
errors will occur in the predicted $J-K$ colors (in whatever
IR photometric system is adopted for the output of the model)
which reproduce the sign and approximate
magnitude of the offset seen
between the model SSP integrated light
IR--IR colors and our measured ones
for Galactic GCs in the bottom panels of Fig.~\ref{figure_model_comp}.
Construction of models, as well as prediction and testing
of integrated colors from them, requires 
careful attention to the details of the calibration of any
stellar or integrated light photometry used in that process.

Only the predicted SSP $(V-K_s)_0 -$ colors of \cite{worthey94} reproduce the
very red colors we find for Galactic GCs at metallicities between
$-0.5$ dex and the Solar value.   It must be emphasized 
that the validity of the models 
cannot be probed at metallicities above Solar from this dataset, as
the sample of well studied Galactic GCs with such high metallicities
is small to non-existent.

For an old (age $\sim$10 Gyr)
single-burst population, an uncertainty in color of 0.10 mag 
corresponds to an error in 
Fe-metallicity of 0.25 dex for $V-K_s$ colors and to 0.7 dex 
for IR--IR colors; the latter is so large
as to render any conclusion regarding the metallicity of a GC useless. 
The apparent mismatch between the SSP model predictions and our
$J-K_s$ colors suggests caution in using IR--IR colors to
determine metallicities (or ages) for  GCs in distant galaxies.

\section{Summary}

We have mosaiced 2MASS images  to derive surface
brightness profiles  in \irfilt\ for 104 Galactic
globular clusters, incorporating algorithms to reduce the impact
of bright field stars.  We fit these with empirical
\cite{king62} profiles,
adopting tidal radii and
cluster center positions from the literature.  This leaves only
the central surface brightness and the core radius as parameters
to be determined.  We show
that the resulting core radii 
for each of these three IR colors
are identical to within the errors.  We therefore set
$r_c$ for each of \irfilt\ to be $r_c(J)$.  We then show
that the $r_c(J)$ for each GC are identical to the core
radii at $V$ in essentially all cases if the uncertainty
for $r_c(V)$ is taken to be 2''.  The only discrepant
cases are core collapsed GCs, where the lower spatial
resolution of 2MASS combined with the small optical
core radii produce smaller measured core radii at $V$ than 
in the IR from 2MASS.  

We derive integrated
light colors $V-J, V-H, V-K_s, J-H$ and $J-K_s$ for these
globular clusters.  We do this by directly integrating
the surface brightness profiles in most cases, equivalet
to slightly censored aperture photometry, as this 
leads to smaller statistical measurement uncertainties than does
integrating the fit King profiles.  
Each color shows a reasonably tight relation
between the dereddened colors and metallicity.  Fits of these
are given for each color.
Linear fits suffice
when the most metal-rich GCs are not considered.  Once
the four GCs in our sample with [Fe/H]$ > -0.2$ dex are included,
a quadratic fit is necessary.  We use our derived $V-K_s$ colors,
combined with total $M_V$ from the database of H96, to find the
luminosity function at $K_s$ of the Galactic globular cluster
system.

The IR--IR colors have very small errors and very low
dispersions about the fits due largely to
the all-sky photometric calibration of the 2MASS survey.  These
errors are consistent with the expected random fluctuations
of the measurements based solely on the measured signal levels,
indicating that other sources of error do not contribute much.
The $V-$IR colors have substantially larger uncertainties
due in part to the lack of an all-sky photometric calibration
for surface brightness measurements at optical wavelengths.
Incorrect choices for reddening, discrepancies in the adopted
position of the center of a cluster, and, for the least populous
GCs, stochastic errors due to the small number of luminous
stars near the tip of the red giant branch, also contribute
to the uncertainties in the $V-$IR colors.   

We find good agreement with measurements of integrated light
colors for a much smaller sample of Galactic globular clusters
by Aaronson, Malkan \& Kleinmann from 1977.  Small constant offsets
between the two datasets of $\sim$0.1 mag in IR--IR colors
are required; we ascribe them to the difficulties
of transforming between the filter and detector system
used in 1977 and the 2MASS system. We find excellent
agreement with the IR--IR colors of \cite{nantais06}, not surprising
since they too use 2MASS images from which to derive their
colors.  But a comparison
at $V-K_s$ of our colors with theirs shows 
very poor agreement in the mean, and with
a very large dispersion; we suspect that they did not
correctly match the optical and IR magnitudes in many cases.  

Our results
provide a calibration for the integrated light of distant
single burst old stellar populations from very low to Solar
metallicities.  We compare our dereddened measured colors
with predictions from several
models of the integrated light of single burst old populations,
bearing in mind that the models have almost
certainly been set up to reproduce the  data of
\cite{frogel83} and \cite{cohen83} for colors of individual red giant
branch stars in Galactic GCs.  While there is reasonable agreement
for $V-K_s$ colors, a $\sim$0.1 mag offset is required in 
$J-K_s$, with the model predictions being redder than our colors.
Until the origin of this problem is understood, 
any determination of [Fe/H] (or age) in old populations 
based on IR--IR colors cannot
be considered valid.
In addition,
some of the models fail to reproduce the behavior of the integrated
light  $V-K_s$ colors of the Galactic globular clusters 
near Solar metallicity.

\acknowledgements

This publication makes use of data products from the Two Micron All Sky
Survey, which is a joint project of the University of Massachusetts and
the IPAC/Caltech, funded by NASA and by NSF.
We thank John Carpenter and Pat C\^ot\'e, who participated in
the initial phase of this work.
J.G.C. is grateful to NSF grant AST-0507219  for partial support.
S.H. was supported by a Summer Undergraduate Research Fellowship,
provided in part through NSF grant AST-0205951 to J.G.C.
S.A.M. acknowledges support from a Spitzer postdoctoral fellowship.
SGD acknowledges partial support from the Ajax Foundation.


\clearpage

%
%
\clearpage
\begin{deluxetable}{lrrr rr rr}
\tablenum{1}
\tablewidth{0pt}
\rotate
\tablecaption{Integrated IR Colors
For GCs With Two Choices for the Position of the Center  
\label{table_center}}
\tablehead{
\colhead{ID} & 
\colhead{RA,Dec (Harris)} &
\colhead{RA,Dec (2MASS)} & \colhead{${\Delta}\theta$} &
\colhead{${\Delta}K_s$\tablenotemark{a}} & 
\colhead{${\Delta}J-K_s$\tablenotemark{a}} &
\colhead{${\Delta}K_s$\tablenotemark{a}} & 
\colhead{${\Delta}J-K_s$\tablenotemark{a}}  \\
\colhead{} &
\colhead{(J2000)} &
\colhead{(J2000)} & \colhead{(arcsec)} &
\colhead{(mag)} & \colhead{(mag)} \\ 
\colhead{} & \colhead{} & \colhead{} & \colhead{} &
\multispan{2}{(King profile int.)} & 
\multispan{2}{(Direct SB prof. int.)}
 }
\startdata 
NGC~5824 &  15 03 58.5 $-33$  04 04 & 15 03 58.30 $-33$  04 07 & 3.9 
 & $0.08$ & $-0.01$ & $-0.01$ & $-0.01$  \\  
NGC~6553 &  18 09 17.6  $-25$ 54 31 & 18  09 17.59 $-25$ 54 38  & 7.0
 &  $-0.11$ &  +0.04 & +0.01 & +0.01 \\
NGC~6715 &  18 55 03.3  $-30$ 28 42 & 18 55 03.50 $-30$ 28 45 & 4.0
 & +0.12 & $-0.04$ & +0.10 & $-0.02$ \\ 
NGC~6838 &   19 53 46.1  +18 46 42 & 19 53 46.10  +18 46 40 & 2.0 
  & +0.10 & $-0.02$ & $-0.06$ & $-0.01$ \\
Pal~6 &  17 43 42.2 $-26$ 13 21 & 17 43 42.29 $-26$ 13 28 & 7.1
 & $-0.18$ & +0.10 & $-0.12$ & +0.02 \\
Pal~8 &   18 41 29.9 $-19$ 49 33 & 18 41 30.09 $-19$ 49 40 & 7.5
 & $-0.10$ & +0.09 & $-0.18$ & +0.15 \\
Terzan~1 &  17 35 47.2 $-30$ 28 54 &  17 35 47.09 $-30$ 28 56 & 2.4 
 & $-0.18$ & $-0.02$ & $-0.09$ & +0.01  \\
Terzan~5 &  17 48 04.9  $-24$ 46 45 & 17 48 05.00 $-24$ 46 49 & 4.2 
  & $-0.01$ & $-0.02$ & $-0.01$ & $-0.01$ \\
~~~~ \\
\multispan{3}{Not treated as 2 GCs} \\
NGC~6426 & 17 44 54.7 +03 10 13 & 17 44 54.4 +03 10 12 & 4.6
    & \nodata &  \nodata \\
NGC~6541 &  18 08 02.2  $-43$ 30 00  & 18  08 02.20 $-43$ 42 20 & 740.0 
 & $>3.0$ & \nodata & \nodata & \nodata  \\    
Terzan~12 &  18 12 15.8  $-22$ 44 31 & 18 12 15.50 $-22$ 44 27 & 5.8 
  & \nodata &  \nodata \\
\enddata
\tablenotetext{a}{Colors with center from H96 -- 
those with new center, 50'' radius aperture used. }
\end{deluxetable}

%
%
\begin{deluxetable}{lrrr rrrr rrrr}
\tablenum{2}
\tabletypesize{\tiny}
\rotate
\tablewidth{0pt}
\tablecaption{Parameters of the King Profile Fits 
For Galactic GCs\tablenotemark{a} \label{table_rcore}}
\tablehead{
\colhead{ID} & \colhead{CC\tablenotemark{b}} & 
\colhead{$r_c(V)$\tablenotemark{c}} & \colhead{$r_c(J)$} &  \colhead{${\sigma}r_c(J)$} &
\colhead{$SB(V)_0$\tablenotemark{d}} & \colhead{$A_0(J)$} &  \colhead{${\sigma}A_0(J)$} &
 \colhead{$A_0(H)$} &  \colhead{${\sigma}A_0(H)$} &
 \colhead{$A_0(K)$} &  \colhead{${\sigma}A_0(K)$} \\
\colhead{} & \colhead{} & \colhead{(arcsec)} & \colhead{(arcsec)} &
\colhead{(arcsec)} & \colhead{(Mag/sq '')} & 
\colhead{(DN/sq '')} & \colhead{(DN/sq '')} & \colhead{(DN/sq '')} & 
\colhead{(DN/sq '')} & \colhead{(DN/sq '')} & \colhead{(DN/sq '')} \\
 }
\startdata 
NGC 362                     & C? & 
  11.4 &   11.9 &    0.8 & 
  14.88 &   1240.9 &    146.3 &   1143.1 &    165.9 &    991.6 &    165.5 \\
 NGC 1261                  &    & 
  23.5 &   27.0 &    1.6 & 
  17.65 &     88.8 &      4.8 &     86.1 &      4.8 &     60.4 &      3.6 \\
NGC 1851                   &    & 
   4.0 &    8.3 &    0.8 & 
  14.15 &   1320.9 &    233.7 &   1251.0 &     99.9 &    922.4 &     63.7 \\
 NGC 1904     & C? & 
   9.6 &   16.6 &    1.7 & 
  16.23 &    218.2 &     30.2 &    219.1 &     17.0 &    152.9 &     11.9 \\
      NGC 2298             &    & 
  20.4 &   27.0 &    2.2 & 
  18.79 &     41.4 &      2.0 &     39.6 &      1.9 &     26.5 &      1.7 \\
    NGC 2419               &    & 
  20.9 &   20.6 &    1.2 & 
  19.83 &     15.1 &      0.6 &     15.9 &      0.5 &     10.9 &      0.4 \\
NGC 2808                   &    & 
  15.8 &   17.4 &    0.8 & 
  15.17 &   1484.8 &    103.5 &   1580.7 &     85.6 &   1159.8 &     49.8 \\
NGC 3201                   &    & 
  83.6 &   97.5 &   12.8 & 
  18.77 &     44.9 &      5.4 &     34.7 &      7.4 &     27.1 &      3.7 \\
NGC 4147                    &    & 
   6.0 &    6.5 &    0.8 & 
  17.63 &     78.6 &     12.2 &     78.9 &      5.2 &     52.7 &      3.8 \\
NGC 4590                    &    & 
  41.7 &   45.1 &    5.4 & 
  18.67 &     30.1 &      3.2 &     35.1 &      4.9 &     16.7 &      1.1 \\
NGC 4833 & \nodata & 
  60.0 &   96.2 &   14.3 & 
  18.45 &     50.1 &      4.0 &     53.3 &      4.2 &     36.2 &      3.4 \\
NGC 5286                    &    & 
  17.4 &   16.8 &    1.5 & 
  16.07 &    659.8 &     81.7 &    711.2 &     52.9 &    506.6 &     40.8 \\
NGC 5634                   &    & 
  12.6 &   11.7 &    1.0 & 
  17.49 &    102.5 &      9.7 &    104.1 &      6.4 &     68.5 &      3.8 \\
NGC 5694                   &    & 
   3.7 &    6.6 &    0.6 & 
  16.34 &    197.0 &     27.1 &    215.2 &     10.9 &    144.2 &      7.3 \\
NGC 5824   &    & 
   3.3 &    3.9 &    0.5 & 
  15.08 &    979.3 &    246.7 &   1127.9 &     87.4 &    691.7 &     38.2 \\
NGC 5904                 &    & 
  23.0 &   29.5 &    1.5 & 
  16.05 &     386.3 &   16.5 &    343.8 &   48.0 &    335.5 &  41.5  \\
NGC 5927                    &    & 
  25.1 &   33.4 &    3.6 & 
  17.45 &    200.1 &     22.0 &    243.1 &     26.0 &    177.3 &     25.2 \\
NGC 5946                    &  C & 
   4.8 &    5.4 &    1.2 & 
  17.42 &    456.1 &    106.6 &    537.9 &     56.6 &    380.8 &     52.8 \\
 NGC 5986                  &    & 
  38.0 &   36.3 &    3.6 & 
  17.56 &    161.4 &     13.6 &    164.8 &     12.4 &    122.1 &      9.4 \\
NGC 6093           &    & 
   8.9 &   10.2 &    0.8 & 
  15.19 &   1038.9 &    130.9 &   1107.9 &     57.3 &    823.0 &     40.1 \\
NGC 6121   &    & 
  38.9 &  112.3 &    5.4 & 
  17.88 &    111.9 &      3.8 &    121.1 &      4.7 &     67.5 &     27.1 \\
NGC 6139                   &    & 
   8.1 &   11.2 &    1.6 & 
  17.30 &    657.4 &    139.0 &    847.2 &     66.7 &    636.0 &     49.0 \\
NGC 6171                   &    & 
  32.3 &   39.4 &    3.7 & 
  18.84 &     54.1 &      5.4 &     65.3 &      5.6 &     45.3 &      4.3 \\
NGC 6205        &    & 
  48.9 &   46.7 &    2.5 & 
  16.80 &    214.4 &     18.0 &    200.6 &     13.9 &    152.9 &      7.1 \\
NGC 6229                    &    & 
   7.9 &    7.3 &    0.4 & 
  16.99 &    266.7 &     23.7 &    283.1 &     12.4 &    198.5 &      8.4 \\
NGC 6218           &    & 
  44.0 &   70.6 &    7.8 & 
  18.17 &     60.8 &      4.4 &     57.1 &      4.9 &     38.6 &      4.6 \\
NGC 6235                    &    & 
  21.4 &   19.2 &    6.1 & 
  18.98 &     52.3 &     15.7 &     60.8 &      8.9 &     44.7 &      6.9 \\
NGC 6254         &    & 
  51.3 &   61.9 &    5.9 & 
  17.69 &    101.5 &      8.7 &     90.9 &     17.7 &     57.5 &     15.0 \\
NGC 6256                   &  C & 
   1.2 &   22.9 &    6.6 & 
  17.89 &     85.7 &     16.2 &    121.7 &     18.5 &     93.6 &     16.9 \\
NGC 6266                   & C? & 
  10.8\tablenotemark{e} &   25.6 &    2.9 & 
  15.35 &   1314.3 &    182.0 &   1586.9 &    110.6 &   1187.8 &     94.8 \\
NGC 6273                   &    & 
  25.7 &   27.2 &    1.6 & 
  16.82 &    489.6 &     37.6 &    524.6 &     27.2 &    366.1 &     24.4 \\
NGC 6284                   &  C & 
   4.2 &    7.9 &    0.7 & 
  16.65 &    383.1 &     35.5 &    464.3 &     31.9 &    313.9 &     25.5 \\
 NGC 6287                  &    & 
  15.8 &   26.3 &    2.1 & 
  18.33 &    127.7 &      7.7 &    153.6 &      7.5 &    116.1 &      5.7 \\
NGC 6293                   &  C & 
   3.0 &   11.1 &    2.0 & 
  16.18 &    393.5 &     95.8 &    413.5 &     43.3 &    289.2 &     35.3 \\
 NGC 6304                  &    & 
  12.6 &   15.0 &    1.2 & 
  17.34 &    414.0 &     41.5 &    510.5 &     36.1 &    363.7 &     34.4 \\
 NGC 6316                  &    & 
  10.0 &   11.7 &    0.7 & 
  17.40 &    550.0 &     26.5 &    733.0 &     32.0 &    549.5 &     29.5 \\
  NGC 6325                 &  C & 
   1.8 &    9.7 &    1.7 & 
  17.56 &    209.4 &     42.1 &    276.4 &     14.1 &    219.1 &     10.5 \\
 NGC 6333         &    & 
  34.7 &   25.0 &    3.3 & 
  17.40 &    279.0 &     36.9 &    312.8 &     23.9 &    218.0 &     19.0 \\
  NGC 6341         &    & 
  14.1 &   16.0 &    1.0 & 
  15.58 &    581.5 &     53.1 &    542.5 &     52.4 &    361.6 &     34.8 \\
 NGC 6342                  &  C & 
   3.0 &    9.3 &    1.0 & 
  17.44 &    185.9 &     23.7 &    227.6 &     11.9 &    156.9 &     15.9 \\
 NGC 6352                  &    & 
  50.1 &   34.1 &   10.3 & 
  18.42 &     56.3 &      9.5 &     57.4 &      6.6 &     41.3 &      5.7 \\
 NGC 6355                  &  C & 
   3.0 &    8.7 &    1.0 & 
  18.05 &    335.1 &     57.2 &    440.9 &     27.1 &    329.7 &     22.6 \\
 NGC 6356                  &    & 
  13.8 &   17.4 &    1.0 & 
  17.09 &    354.3 &     22.1 &    436.9 &     18.3 &    317.4 &     14.8 \\
  NGC 6380                 & C? & 
  20.4 &   21.7 &    2.1 & 
  19.96 &    183.6 &     14.1 &    293.3 &     15.6 &    237.5 &     21.1 \\
NGC 6388                   &    & 
   7.4 &   12.7 &    1.5 & 
  14.55 &   3037.5 &    614.1 &   3719.9 &    205.2 &   2796.2 &    161.0 \\
NGC 6402          &    & 
  50.1 &   47.4 &    5.1 & 
  18.41 &    144.7 &     15.0 &    169.7 &     11.9 &    142.0 &     22.7 \\
NGC 6397                   &  C & 
   3.0 &   61.5 &    9.3 & 
  15.65 &    129.6 &     12.6 &    121.3 &     12.8 &     81.4 &      8.9 \\
NGC 6401                   &    & 
  14.8 &   11.4 &    1.1 & 
  18.67 &    248.6 &     23.3 &    323.5 &     30.8 &    230.2 &     29.9 \\
 NGC 6426  &    & 
  15.8 &    2.0 &    1.5 & 
  20.37 &    271.3 &    361.8 &    407.5 &    122.0 &    297.2 &    101.8 \\
NGC 6440                    &    & 
   7.6 &   10.8 &    1.2 & 
  17.02 &   1792.8 &    259.4 &   2818.0 &    173.4 &   2332.7 &    153.9 \\
NGC 6441                    &    & 
   6.8 &   10.6 &    0.4 & 
  14.99 &   2765.7 &    138.4 &   3617.5 &    100.7 &   2737.8 &     72.0 \\
NGC 6453                    &  C & 
   4.2 &    7.6 &    1.2 & 
  17.35 &    394.4 &     55.0 &    475.6 &     33.1 &    336.0 &     39.6 \\
NGC 6496                     &    & 
  63.1 &  115.9 &   30.1 & 
  20.10 &     37.2 &      9.8 &     37.2 &      3.8 &     26.1 &      2.7 \\
NGC 6517                    &    & 
   3.7 &    8.7 &    1.9 & 
  17.77 &    547.8 &    155.9 &    753.1 &     88.5 &    602.7 &     67.8 \\
       NGC 6522            &  C & 
   3.0 &   20.2 &    2.8 & 
  16.14 &    259.3 &     26.7 &    315.8 &     27.0 &    226.4 &     21.7 \\
       NGC 6535            &    & 
  25.1 &   11.6 &    3.1 & 
  20.22 &     23.8 &      6.5 &     25.9 &      3.3 &     16.5 &      2.7 \\
       NGC 6539            &    & 
  32.3 &   32.9 &    3.4 & 
  19.31 &     90.1 &      9.1 &    124.8 &      8.8 &    110.5 &      8.7 \\
       NGC 6540   & \nodata & 
   1.8 &    1.4 &    1.6 & 
  16.40 &   1321.0 &   2909.1 &   1335.6 &    269.4 &    767.2 &    223.0 \\
      NGC 6541 & C? & 
   7.2 &   16.7 &    1.3 & 
  15.58 &    575.1 &     48.8 &    586.4 &     56.0 &    411.0 &     48.3 \\
       NGC 6544            & C? & 
  13.2 &   32.0 &    7.2 & 
  17.31 &    588.0 &     78.6 &    694.5 &     84.5 &    522.7 &     63.0 \\
      NGC 6553             &    & 
  33.1 &   32.9 &    4.4 & 
  18.15 &    614.1 &     88.1 &    938.7 &    139.7 &    802.6 &    158.8 \\
     NGC 6558              &  C & 
   1.8 &    4.2 &    0.5 & 
  17.08 &    307.6 &     47.4 &    325.5 &     27.6 &    206.7 &     28.9 \\
     NGC 6569              &    & 
  22.4 &   19.7 &    3.7 & 
  18.08 &    244.9 &     45.0 &    304.9 &     41.4 &    215.2 &     32.4 \\
     NGC 6584              &    & 
  35.5 &   26.1 &    3.1 & 
  17.79 &     62.8 &      8.8 &     67.5 &      7.3 &     40.6 &      3.8 \\
    NGC 6624               &  C & 
   3.6 &    7.0 &    0.7 & 
  15.42 &   1162.1 &    150.7 &   1338.5 &    143.2 &   1024.5 &     97.4 \\
    NGC 6626       &    & 
  14.5 &   18.2 &    1.7 & 
  16.08 &    825.7 &     98.3 &    968.4 &     60.2 &    687.3 &     38.1 \\
    NGC 6637         &    & 
  20.4 &   19.2 &    2.3 & 
  16.83 &    436.6 &     74.6 &    525.4 &     37.0 &    367.6 &     30.4 \\
    NGC 6638               &    & 
  15.8 &   12.0 &    1.0 & 
  17.27 &    309.8 &     25.8 &    343.8 &     28.1 &    251.8 &     27.3 \\
    NGC 6642               & C? & 
   6.0 &    3.3 &    1.6 & 
  16.68 &    940.7 &    799.6 &   1074.4 &    134.4 &    745.2 &     93.0 \\
    NGC 6652               &    & 
   4.3 &    8.2 &    1.4 & 
  16.31 &    345.5 &     72.6 &    366.1 &     37.3 &    245.7 &     29.0 \\
   NGC 6656          &    & 
  85.1 &  121.8 &   15.3 & 
  17.32 &    184.7 &     13.9 &    172.6 &     16.3 &    127.7 &     11.3 \\
    NGC 6681         &  C & 
   1.8 &    6.2 &    1.3 & 
  15.28 &    607.9 &    189.0 &    624.3 &     56.1 &    420.4 &     41.3 \\
   NGC 6712                &    & 
  56.2 &   50.7 &    8.4 & 
  18.65 &    111.7 &     12.2 &    128.8 &     12.6 &     94.1 &      9.2 \\
  NGC 6715  &    & 
   6.5 &    7.4 &    0.3 & 
  14.82 &   1521.6 &     87.6 &   1697.2 &     57.6 &   1179.3 &     52.6 \\
 NGC 6717           & C? & 
   4.8 &    6.3 &    0.9 & 
  16.48 &    320.1 &     68.8 &    348.2 &     22.2 &    239.5 &     14.9 \\
 NGC 6723                  &    & 
  56.2 &   47.6 &    2.7 & 
  17.92 &    106.3 &      6.9 &    108.9 &      5.0 &     75.0 &      3.7 \\
 NGC 6749  & \nodata & 
  46.2 &   56.1 &    4.1 & 
  21.54 &    101.7 &      3.4 &    153.4 &      4.8 &    124.2 &      4.3 \\
 NGC 6760                  &    & 
  20.0 &   17.3 &    2.8 & 
  18.79 &    349.9 &     60.3 &    517.8 &     56.0 &    406.5 &     56.0 \\
 NGC 6779      &    & 
  21.9 &   33.2 &    3.4 & 
  18.06 &     88.7 &      8.2 &     96.0 &      7.1 &     62.7 &      4.9 \\
 NGC 6809              &    & 
 169.7 &  144.9 &   10.3 & 
  19.13 &     44.8 &      2.4 &     47.3 &      3.1 &     30.5 &      2.0 \\
NGC 6838   & \nodata & 
  37.8 &   48.6 &    9.0 & 
  19.22 &     60.8 &      7.6 &     65.5 &      7.1 &     45.8 &      4.4 \\
NGC 6864                   &    & 
   5.8 &    7.2 &    0.6 & 
  15.55 &    793.2 &    109.9 &    869.3 &     34.1 &    619.4 &     33.7 \\
NGC 6934                   &    & 
  14.8 &   13.7 &    0.6 & 
  17.26 &    164.3 &      8.5 &    162.9 &      7.0 &    115.5 &      3.3 \\
NGC 6981            &    & 
  32.3 &   26.7 &    1.2 & 
  18.90 &     32.1 &      1.3 &     34.0 &      2.8 &     22.8 &      2.2 \\
NGC 7006                   &    & 
  14.5 &    8.4 &    0.8 & 
  18.50 &     73.5 &      8.4 &     76.9 &      4.2 &     52.4 &      2.2 \\
NGC 7089   &    & 
  20.4 &   19.7 &    2.0  & 
  15.92 &  504.2 & 32.5 &   522.0 & 15.0 & 315.5 & 33.0  \\
NGC 7099  &  C & 
   3.6 &   14.7 &    1.2 & 
  15.28 &    243.9 &     25.0 &    206.9 &     36.9 &    131.2 &     32.1 \\
Pal 2   &    & 
   14.4 &   12.5 &    1.0 & 
  19.39 &    14.4 &      3.3 &     92.0 &      3.6 &     75.0 &      3.3 \\
Pal 6    &    & 
  39.8 &   29.8 &    7.9 & 
  21.58 &    115.6 &     20.3 &    195.3 &     32.3 &    170.7 &     32.0 \\
Pal 7  &    & 
  64.6 &   98.2 &   15.2 & 
  21.66 &     27.7 &      2.2 &     39.9 &      3.5 &     22.8 &      7.0 \\
    Pal 8   &    & 
  24.0 &   26.9 &    5.9 & 
  19.83 &     19.4 &      1.6 &     22.5 &      1.7 &     16.5 &      1.4 \\
 Terzan 1   &  C & 
   2.4 &    9.8 &    0.9 & 
  25.09 &    903.7 &    118.7 &   1725.4 &     95.5 &   1589.4 &    102.2 \\
Terzan 2                   &  C & 
   1.8 &    5.7 &    0.8 & 
  21.58 &    505.5 &    104.2 &    922.1 &     75.1 &    801.6 &     68.0 \\
Terzan 5    &    & 
  14.5 &    6.9 &    0.5 & 
  20.33 &   3208.7 &    257.7 &   7636.8 &    328.1 &   8107.6 &    365.3 \\
Terzan 6                    &  C & 
   3.0 &    7.8 &    2.1 & 
  20.76 &    170.4 &     33.6 &    386.4 &     51.7 &    403.2 &     65.9 \\
Terzan 9                    &  C & 
   1.8 &    7.2 &    1.8 & 
  23.21 &    539.8 &    148.8 &    989.5 &    129.3 &    869.4 &    122.1 \\
 Terzan 12  & \nodata & 
   \nodata &   24.0 &    8.6 & 
  \nodata &     75.8 &      8.5 &    148.4 &      9.8 &    132.0 &      7.8 \\
Djorg 1 & \nodata & 
  19.2 &    8.1 &    2.3 & 
  23.10 &     64.7 &     17.1 &    103.5 &     13.3 &     90.8 &     10.9 \\
     HP 1  &  C & 
   \nodata &   15.7 &    3.7 & 
  \nodata &    126.1 &     28.3 &    178.9 &     28.4 &    123.2 &     21.6 \\
Liller 1    & \nodata & 
   \nodata &    5.2 &    0.5 & 
  \nodata &    858.4 &    130.1 &    858.4 &    130.1 &   3337.1 &    220.0 \\
 Ton 2   &  \nodata &
  32.4 &   24.8 &    4.3 & 
  22.16 &     83.9 &      6.1 &    132.4 &      7.9 &    106.0 &      8.0 \\
UKS 1   & \nodata & 
   9.0 &   13.5 &    1.1 & 
  25.52 &    178.9 &     12.7 &    547.7 &     21.1 &    625.6 &     25.5 \\
\enddata
\tablenotetext{a}{Values are given as measured, without reddening corrections}.
\tablenotetext{b}{C denotes known core-collapsed GC, ``c?''
indicates known probable core-collapsed GC.}
\tablenotetext{c}{$r_c(V)$ preferentially from TKD95, or from H96.  
$r_c(V)$ from H96 for all probable core-collapsed GCs.}
\tablenotetext{d}{$SB(V)_0$ from H96.}
\tablenotetext{e}{\cite{ngc6266} gives $r_c(V) = 19$'' for NGC~6266
and find that it is not a core-collapsed GC.}
\end{deluxetable}


\clearpage

\clearpage

%
%
\begin{deluxetable}{lrrrc  rrrrr r}
\tablenum{3}
\tabletypesize{\tiny}
\tablewidth{0pt}
\tablecaption{Reddening Corrected Integrated Light IR Colors for Galactic GCs -- 50'' Radius Aperture
\label{table_colors50}}
\tablehead{
\colhead{ID} & \colhead{Class} &  
\colhead{$E(B-V)$} & \colhead{[Fe/H]} & \colhead{[Fe/H]} &
\colhead{$V-K_s$} &  
\colhead{$J-H$} & \colhead{${\sigma}(J-H)$} &
\colhead{$J-K_s$} & \colhead{${\sigma}(J-K_s)$} 
& \colhead{$J^{50}$} \\
\colhead{} & \colhead{\tablenotemark{a}} & 
\colhead{(mag)} & \colhead{(dex)} & \colhead{Code\tablenotemark{b}} &
\colhead{(mag)} & \colhead{(mag)} & \colhead{(mag)}
 & \colhead{(mag)} & \colhead{(mag)} & \colhead{(mag)} \\
}
\startdata
 NGC 104   & BB &     0.04 &    $-$0.78 &  KI & 2.60 &     0.55 &     0.03 &     0.74 &     0.03 &     3.83  \\ 
 NGC 362   & FF &     0.05 &    $-$1.21 &  KI & 2.53 &     0.52 &     0.03 &     0.64 &     0.03 &     5.32  \\ 
 NGC 1261  & BB &     0.10 &    $-$1.19 &  KI & 2.05 &     0.44 &     0.05 &     0.50 &     0.06 &     7.31  \\ 
 NGC 1851  & BB &     0.02 &    $-$1.12 &  KI & 2.66 &     0.53 &     0.03 &     0.65 &     0.03 &     5.90  \\ 
 NGC 1904  & BB &     0.01 &    $-$1.59 &  KI & 2.08 &     0.49 &     0.04 &     0.58 &     0.05 &     7.16  \\ 
 NGC 2298  & BB &     0.14 &    $-$2.04 &  KI & 2.39 &     0.38 &     0.11 &     0.52 &     0.13 &     8.17  \\ 
 NGC 2419  & BB &     0.11 &    $-$2.12 &  KI & 2.20 &     0.41 &     0.32 &     0.48 &     0.39 &     9.48  \\ 
 NGC 2808  & BB &     0.22 &    $-$1.23 &  ZW,H96 & 2.50 &     0.51 &     0.25 &     0.62 &     0.03 &     4.64  \\ 
 NGC 3201  & FF &     0.23 &    $-$1.40 &  KI & 2.09 &     0.44 &     0.05 &     0.51 &     0.07 &     7.02  \\ 
 NGC 4147  & BB &     0.02 &    $-$1.75 &  KI & 1.88 &     0.41 &     0.25 &     0.50 &     0.31 &     9.29  \\ 
 NGC 4590  & BB &     0.05 &    $-$2.43 &  KI & 1.94 &     0.43 &     0.09 &     0.45 &     0.13 &     8.00  \\ 
 NGC 4833  & BB &     0.32 &    $-$2.04 &  KI & 2.14 &     0.39 &     0.04 &     0.43 &     0.05 &     6.46  \\ 
 NGC 5272  & BB &     0.01 &    $-$1.36 &  CM & 2.23 &     0.52 &     0.03 &     0.61 &     0.04 &     5.96  \\ 
 NGC 5286  & BB &     0.24 &    $-$1.65 &  KI & 2.26 &     0.49 &     0.03 &     0.55 &     0.03 &     5.70  \\ 
 NGC 5634  & BB &     0.05 &    $-$1.88 &  ZW,H96 & 1.85 &     0.36 &     0.11 &     0.48 &     0.13 &     8.22  \\ 
 NGC 5694  & BB &     0.09 &    $-$2.08 &  KI & 2.26 &     0.51 &     0.13 &     0.56 &     0.16 &     8.52  \\ 
 NGC 5824  & BB &     0.13 &    $-$1.85 &  ZW,H96 & 2.21 &     0.46 &     0.05 &     0.54 &     0.06 &     7.33  \\ 
 NGC 5904  & FF &     0.03 &    $-$1.25 &  KI & 2.19 &     0.49 &     0.03 &     0.61 &     0.03 &     5.67  \\ 
 NGC 5927  & AF &     0.45 &    $-$0.37 &  ZW,H96 & 3.17 &     0.58 &     0.03 &     0.74 &     0.03 &     5.67  \\ 
 NGC 5946  & AF &     0.54 &    $-$1.38 &  ZW,H96 & 1.98 &     0.49 &     0.04 &     0.55 &     0.04 &     6.67  \\ 
 NGC 5986  & BB &     0.28 &    $-$1.56 &  KI & 2.15 &     0.47 &     0.03 &     0.56 &     0.03 &     6.13  \\ 
 NGC 6093  & BB &     0.18 &    $-$1.72 &  KI & 2.23 &     0.50 &     0.03 &     0.59 &     0.03 &     5.79  \\ 
 NGC 6121  & AA &     0.36 &    $-$1.15 &  KI & 2.24 &     0.44 &     0.03 &     0.52 &     0.03 &     5.77  \\ 
 NGC 6139  & AB &     0.75 &    $-$1.68 &  ZW,H96 & 2.49 &     0.52 &     0.03 &     0.55 &     0.03 &     5.72  \\ 
 NGC 6171  & BB &     0.33 &    $-$1.02 &  KI & 2.64 &     0.53 &     0.04 &     0.60 &     0.05 &     6.86  \\ 
 NGC 6205  & BB &     0.02 &    $-$1.53 &  KI & 2.11 &     0.46 &     0.03 &     0.54 &     0.03 &     5.93  \\ 
 NGC 6218  & FF &     0.19 &    $-$1.26 &  KI & 1.92 &     0.37 &     0.04 &     0.46 &     0.06 &     6.99  \\ 
 NGC 6229  & BB &     0.01 &    $-$1.43 &  ZW,H96 & 2.43 &     0.53 &     0.08 &     0.63 &     0.10 &     8.00  \\ 
 NGC 6235  & FF &     0.36 &    $-$1.32 &  KI & 1.95 &     0.51 &     0.11 &     0.57 &     0.13 &     8.08  \\ 
 NGC 6254  & AA &     0.28 &    $-$1.43 &  KI & 2.19 &     0.47 &     0.03 &     0.54 &     0.03 &     5.91  \\ 
 NGC 6256  & AA &     1.03 &    $-$0.70 &  H96 & 3.00 &     0.61 &     0.04 &     0.63 &     0.05 &     6.40  \\ 
 NGC 6266  & AB &     0.47 &    $-$1.12 &  KI & 2.33 &     0.54 &     0.03 &     0.62 &     0.03 &     4.26  \\ 
 NGC 6273  & AB &     0.41 &    $-$1.79 &  KI & 2.46 &     0.48 &     0.03 &     0.54 &     0.03 &     5.11  \\ 
 NGC 6284  & BB &     0.28 &    $-$1.32 &  ZW,H96 & 2.47 &     0.44 &     0.03 &     0.52 &     0.04 &     6.73  \\ 
 NGC 6287  & AB &     0.60 &    $-$2.05 &  ZW & 2.56 &     0.53 &     0.03 &     0.61 &     0.03 &     6.46  \\ 
 NGC 6293  & AF &     0.41 &    $-$1.92 &  ZW & 2.21 &     0.43 &     0.03 &     0.47 &     0.03 &     6.20  \\ 
 NGC 6304  & AB &     0.53 &    $-$0.59 &  ZW & 2.69 &     0.59 &     0.03 &     0.71 &     0.03 &     5.63  \\ 
 NGC 6316  & AB &     0.51 &    $-$0.55 &  ZW,H96 & 3.19 &     0.65 &     0.03 &     0.77 &     0.03 &     5.92  \\ 
 NGC 6325  & AB &     0.89 &    $-$1.17 &  ZW,H96 & 2.41 &     0.51 &     0.05 &     0.57 &     0.05 &     6.71  \\ 
 NGC 6333  & BB &     0.38 &    $-$1.75 &  ZW,H96 &   2.17 &     0.47 &     0.03 &     0.53 &     0.03 &     5.90  \\ 
 NGC 6341  & BB &     0.02 &    $-$2.38 &  CMa & 2.06 &     0.45 &     0.03 &     0.53 &     0.03 &     5.88  \\ 
 NGC 6342  & AF &     0.46 &    $-$0.65 &  ZW,H96 & 2.53 &     0.55 &     0.05 &     0.62 &     0.06 &     7.14  \\ 
 NGC 6352  & FF &     0.21 &    $-$0.69 &  KI & 1.98 &     0.48 &     0.06 &     0.59 &     0.07 &     7.25  \\ 
 NGC 6355  & AB &     0.75 &    $-$1.50 &  ZW & 2.66 &     0.58 &     0.03 &     0.64 &     0.03 &     6.36  \\ 
 NGC 6356  & BB &     0.28 &    $-$0.50 &  ZW,H96 &  2.65 &     0.61 &     0.03 &     0.70 &     0.03 &     6.22  \\ 
 NGC 6380  & AA &     1.17 &    $-$0.50 &  Z,H96 & 3.15 &     0.63 &     0.03 &     0.69 &     0.03 &     5.51  \\ 
 NGC 6388  & BB &     0.37 &    $-$0.60 &  ZW,H96 & 2.65 &     0.62 &     0.03 &     0.75 &     0.03 &     4.30  \\ 
 NGC 6397  & FF &     0.18 &    $-$2.11 &  KI & 1.94 &     0.34 &     0.03 &     0.41 &     0.03 &     6.05  \\ 
 NGC 6401  & AF &     0.72 &    $-$0.98 &  ZW,H96 & 2.89 &     0.57 &     0.03 &     0.64 &     0.04 &     6.22  \\ 
 NGC 6402  & AF &     0.60 &    $-$1.39 &  ZW & 2.34 &     0.48 &     0.03 &     0.53 &     0.03 &     5.56  \\ 
 NGC 6426  & AA &     0.36 &    $-$2.26 &  ZW,H96 &   2.66 &     0.57 &     0.27 &     0.58 &     0.31 &     9.17  \\ 
 NGC 6440  & AA &     1.07 &    $-$0.34 &  ZW,H96 &   3.06 &     0.66 &     0.03 &     0.74 &     0.03 &     4.41  \\ 
 NGC 6441  & AB &     0.47 &    $-$0.53 &  ZW,H96 &   2.68 &     0.62 &     0.03 &     0.74 &     0.03 &     4.47  \\ 
 NGC 6453  & AF &     0.66 &    $-$1.53 &  ZW &   2.61 &     0.49 &     0.03 &     0.58 &     0.03 &     6.23  \\ 
 NGC 6496  & FF &     0.15 &    $-$0.69 &  KI & 3.03 &     0.48 &     0.12 &     0.56 &     0.14 &     8.24  \\ 
 NGC 6517  & AA &     1.08 &    $-$1.37 &  ZW,H96 & 3.22 &     0.48 &     0.03 &     0.49 &     0.03 &     6.02  \\ 
 NGC 6522  & AB &     0.48 &    $-$1.36 &  KI & 2.50 &     0.57 &     0.04 &     0.65 &     0.04 &     5.84  \\ 
 NGC 6528  & AB &     0.54 &     0.07 & Car &  4.04 &     0.65 &     0.03 &     0.78 &     0.03 &     5.50  \\ 
 NGC 6535  & FF &     0.34 &    $-$1.76 &  KI & 2.11 &     0.45 &     0.22 &     0.43 &     0.28 &     8.90  \\ 
 NGC 6539  & AB &     0.97 &    $-$0.66 &  ZW & 2.79 &     0.52 &     0.04 &     0.64 &     0.04 &     6.17  \\ 
 NGC 6540  & AA &     0.60 &    $-$1.20 &  H96 & 3.95 &     0.50 &     0.04 &     0.55 &     0.04 &     6.63  \\ 
 NGC 6541  & FF &     0.14 &    $-$1.78 &  KI & 2.33 &     0.41 &     0.13 &     0.42 &     0.15 &     5.66  \\ 
 NGC 6544  & AF &     0.73 &    $-$1.35 & KI &  2.40 &     0.49 &     0.03 &     0.59 &     0.03 &     5.34  \\ 
 NGC 6553  & AF &     0.63 &    $-$0.06 & CohCar & 3.73 &     0.69 &     0.03 &     0.87 &     0.03 &     4.53  \\ 
 NGC 6558  & AF &     0.44 &    $-$1.44 &  ZW & 2.35 &     0.49 &     0.06 &     0.54 &     0.07 &     7.29  \\ 
 NGC 6569  & AF &     0.55 &    $-$0.86 &  ZW & 2.57 &     0.59 &     0.03 &     0.65 &     0.03 &     6.03  \\ 
 NGC 6584  & BB &     0.10 &    $-$1.49 &  ZW,H96 & 2.33 &     0.45 &     0.08 &     0.51 &     0.09 &     7.78  \\ 
 NGC 6624  & BB &     0.28 &    $-$0.69 &  KI & 2.82 &     0.60 &     0.03 &     0.73 &     0.03 &     5.85  \\ 
 NGC 6626  & AB &     0.40 &    $-$1.12 &  KI & 2.43 &     0.52 &     0.03 &     0.60 &     0.03 &     5.02  \\ 
 NGC 6637  & BB &     0.16 &    $-$0.80 &  KI &   2.99 &     0.61 &     0.03 &     0.75 &     0.03 &     5.85  \\ 
 NGC 6638  & AF &     0.40 &    $-$0.92 &  KI &   2.36 &     0.53 &     0.03 &     0.61 &     0.03 &     6.44  \\ 
 NGC 6642  & AF &     0.41 &    $-$1.35 & ZW,H96 &  2.67 &     0.60 &     0.03 &     0.65 &     0.04 &     6.79  \\ 
 NGC 6652  & FF &     0.09 &    $-$0.69 & ZW,H96 & 2.51 &     0.52 &     0.05 &     0.60 &     0.06 &     7.37  \\ 
 NGC 6656  & BB &     0.32 &    $-$1.64 & ZW,H96 & 2.37 &     0.41 &     0.03 &     0.50 &     0.03 &     5.03  \\ 
 NGC 6681  & BB &     0.07 &    $-$1.54 & KI &  2.04 &     0.45 &     0.04 &     0.53 &     0.05 &     7.09  \\ 
 NGC 6712  & AB &     0.45 &    $-$1.02 & KI &    2.26 &     0.44 &     0.03 &     0.50 &     0.04 &     6.43  \\ 
 NGC 6715  & BB &     0.15 &    $-$1.41 & KI &    2.40 &     0.55 &     0.03 &     0.66 &     0.03 &     5.92  \\ 
 NGC 6717  & BB &     0.22 &    $-$1.21 & KI &    2.60 &     0.39 &     0.05 &     0.49 &     0.06 &     7.33  \\ 
 NGC 6723  & BB &     0.05 &    $-$1.03 & KI &    2.42 &     0.49 &     0.03 &     0.57 &     0.04 &     6.74  \\ 
 NGC 6749  & AA &     1.50 &    $-$1.60 & H96 &  2.86 &     0.48 &     0.04 &     0.44 &     0.03 &     5.62  \\ 
 NGC 6760  & AF &     0.77 &    $-$0.52 & ZW &   3.06 &     0.63 &     0.03 &     0.72 &     0.03 &     5.59  \\ 
 NGC 6779  & BB &     0.20 &    $-$1.94 & ZW &   2.11 &     0.47 &     0.04 &     0.53 &     0.05 &     7.11  \\ 
 NGC 6809  & BB &     0.08 &    $-$1.82 & KI &  2.43 &     0.38 &     0.07 &     0.44 &     0.09 &     7.62  \\ 
 NGC 6838  & FF &     0.25 &    $-$0.73 & KI &  2.72 &     0.47 &     0.06 &     0.53 &     0.07 &     7.32  \\ 
 NGC 6864  & BB &     0.16 &    $-$1.16 & ZW,H96 &  2.36 &     0.52 &     0.03 &     0.61 &     0.03 &     6.73  \\ 
 NGC 6934  & BB &     0.10 &    $-$1.54 & ZW &  2.03 &     0.46 &     0.06 &     0.55 &     0.07 &     7.50  \\ 
 NGC 6981  & FF &     0.05 &    $-$1.36 & KI &  1.94 &     0.44 &     0.14 &     0.53 &     0.17 &     8.51  \\ 
 NGC 7006  & BB &     0.05 &    $-$1.63 & ZW,H96 &  2.34 &     0.33 &     0.23 &     0.53 &     0.26 &     9.09  \\ 
 NGC 7089  & FF &     0.06 &    $-$1.51 & KI &  1.63 &     0.47 &     0.03 &     0.57 &     0.03 &     6.18  \\ 
 NGC 7099  & AA &     0.03 &    $-$2.32 & KI &  2.09 &     0.41 &     0.03 &     0.49 &     0.04 &     6.84  \\ 
 IC 1276   & AA &     1.08 &    $-$0.73 & H96 &    3.09 &     0.62 &     0.08 &     0.67 &     0.09 &     6.54  \\ 
 Pal 2     & AA &     1.24 &    $-$1.30 & H96 &  \nodata &     0.56 &     0.13 &     0.56 &     0.14 &     7.65  \\ 
 Pal 6     & AA &     1.46 &    $-$1.09 & Z,H96 &    3.51 &     0.67 &     0.09 &     0.72 &     0.10 &     5.31  \\
 Pal 8     & FF &     0.32 &    $-$0.48 & ZW &  3.00 &     0.54 &     0.13 &     0.62 &     0.14 &     7.86  \\ 
 Terzan 1  & AA &     2.28 &    $-$1.30 & AZ,H96 &  \nodata &     0.52 &     0.03 &     0.44 &     0.03 &     3.88  \\ 
 Terzan 2  & AA &     1.57 &    $-$0.40 & AZ,H96 &  \nodata &     0.58 &     0.03 &     0.63 &     0.03 &     5.54  \\ 
 Terzan 5  & AA &     2.15 &     0.00 & AZ,H96 &    3.63 &     0.75 &     0.03 &     0.80 &     0.03 &     3.26  \\ 
 Terzan 6  & AA &     2.14 &    $-$0.50 &  AZ,H96 & \nodata &     0.74 &     0.06 &     0.77 &     0.05 &     5.39  \\ 
 Terzan 9  & AA &     1.87 &    $-$2.00 & AZ,H96 &  \nodata &     0.57 &     0.03 &     0.55 &     0.03 &     4.96  \\
 Terzan12  & AA &     2.06 &    $-$0.50 &  H96 & \nodata &     0.73 &     0.06 &     0.72 &     0.06 &     5.52  \\ 
 HP 1      & AF &     0.74 &    $-$1.55 & AZ,H96 &  \nodata &     0.74 &     0.09 &     0.85 &     0.10 &     6.09  \\ 
 UKS 1     & AA &     3.09 &    $-$0.50 & ZW,H96 & \nodata &     0.68 &     0.03 &     0.61 &     0.03 &     4.25  \\ 
 Djorg 1   & AA &     1.44 &    $-$2.00 & H96 & 4.72\tablenotemark{c} &    
  0.65 &     0.13 &     0.77 &     0.12 &     6.37  \\ 
 Ton 2     & AA &     1.24 &    $-$0.50 &  H96 & 4.17\tablenotemark{c} &  
       0.62 &     0.05 &     0.66 &     0.05 &     5.95  \\ 
 Liller 1  & AA &     3.06 &     0.22 &  AZ,H96 & \nodata &     0.77 &     0.03 &     0.78 &     0.03 &     4.18  \\   
\enddata
\tablenotetext{a}{Class($V-K_s$), Class(IR), B = Best, F = Fair and not Best, 
A = All, i.e. not B or F}
\tablenotetext{b}{[Fe/H] sources: KI = \cite{kraft03}, AZ = \cite{arm88}, Z = \cite{zinn85},
H96 = \cite{harris96} and references therein, Coh = \cite{cohen99},  
Car = \cite{carretta01}, CM = \cite{cohen05}, CMa = Cohen \& Melendez, in preparation.}
\tablenotetext{c}{These GCs appear to have problems in the $V$ SB zero point or
substantial errors in $E(B-V)$.}
\end{deluxetable}


\clearpage

%
%
\begin{deluxetable}{lrrr}
\tablenum{4}
\tablewidth{0pt}
\tablecaption{Numbers of Galactic GCs in Our Samples
\label{table_n_samp}}
\tablehead{
\colhead{Group} & \colhead{$E(B-V)$(Below)} & \colhead{SNR($K_s$)(min)\tablenotemark{a}} 
& \colhead{Number} \\
\colhead{} & \colhead{(mag)} & \colhead{} & \colhead{} \\
 }
\startdata 
$J-K_s$ \\
Best & 1.0 & 10 & 52 \\
Fair & 1.0 & 5 & 82 \\
All & \nodata & \nodata & \ntotal\tablenotemark{b} \\
~~~ \\
$V-K_s$ \\
Best & 0.40 & 10 & 38 \\
Fair & 0.40 & 5 & 53 \\
All & \nodata & \nodata & 96\tablenotemark{bc} \\
\enddata
\tablenotetext{a}{SNR determined from fit King profile surface brightness 
evaluated in central 5'' of GC.  Actual SNR from pseudo-aperture 
photometry is much higher.}
\tablenotetext{b}{This includes 47 Tuc, with IR data from the
2MASS Large Galaxy Atlas.}
\tablenotetext{c}{Nine of the sample GCs have no accurate $V$ 
surface brightness profile.}
\end{deluxetable}

\clearpage

%
%
\begin{deluxetable}{lrrrr r}
\tablenum{5}
\tablewidth{0pt}
\tablecaption{Aaronson, Malkan \& Kleinmann 1978
Integrated Light Photometry of Galactic GCs 
\label{table_acmm}}
\tablehead{
\colhead{ID} & \colhead{$V-K$\tablenotemark{a}} & 
\colhead{$J-H$\tablenotemark{a}} & \colhead{$H-K_s$\tablenotemark{a}} &
\colhead{CO Index\tablenotemark{a}} & \colhead{$H_2O$ Index\tablenotemark{a}} \\
\colhead{} & \colhead{(mag -- Obs.)} & \colhead{(mag -- Obs)} &
\colhead{(mag -- Obs)} &
\colhead{(mag -- Dered.)} & \colhead{(mag  -- Dered.)} \\
}
\startdata
Low Red Calibs \\
NGC~5024 &   2.19 &   0.52 & 0.11 &  0.016&   0.063 \\ 
NGC~5272 &   2.22 &   0.55 & 0.08 &  0.021&   0.024 \\ 
NGC~5904 &   2.33 &   0.60 & 0.10 &  0.044&   0.013 \\ 
NGC~6205 &   2.62 &   0.59 & 0.10 &  0.031&   0.034 \\ 
NGC~6254 &   3.10\tablenotemark{b} &   0.65 & 0.13 &   0.022&   0.044 \\ 
NGC~6341 &   2.20 &   0.48 & 0.10 & $-0.006$&   0.022 \\ 
NGC~6838 &   3.70\tablenotemark{b} &   0.72 & 0.17 &  0.075&   0.045 \\ 
NGC~7006 &   2.37 &   0.53 & 0.09 & \nodata & \nodata \\ 
NGC~7078 &   2.16 &   0.47 & 0.11 & $-0.006$ &   0.017 \\ 
NGC~7089 &   2.38 &   0.53 & 0.09 &  0.024&   0.031 \\ 
NGC~7099 &   2.09 &   0.49 & 0.09 &  0.007&   0.039 \\ 
~ \\
High Red Calibs \\
NGC~6121 &   3.54 &   0.75 & 0.17 &   0.065 &   0.015 \\
NGC~6171 &   4.03\tablenotemark{b} & 0.78 & 0.19 &   0.089 &   0.052 \\
NGC~6656 &   3.16 &  0.69 & 0.14 &   0.038 &   0.040 \\
~ \\
Low Red \\
NGC~1904 &   2.24\tablenotemark{c} &   0.52 & 0.07 & \nodata & \nodata \\ 
NGC~2419 &   2.17 &   0.54 & 0.04 & \nodata & \nodata \\ 
NGC~5634 &   2.44\tablenotemark{c} &   0.53 & 0.11 &   0.054&   0.051 \\ 
NGC~6093 &   2.92 &   0.61 & 0.11 &   0.011&   0.042 \\ 
NGC~6218 &   2.68 &   0.62 & 0.11 &  0.041&   0.092 \\ 
NGC~6229 &   2.50 &   0.58 & 0.08 & \nodata & \nodata \\ 
NGC~6356 &   3.65 &   0.76 & 0.19 &   0.063&   0.048 \\ 
NGC~6637 &   3.23 &   0.80 & 0.17 &  0.073&   0.064 \\ 
NGC~6715 &   2.78 &   0.62 & 0.14 & \nodata & \nodata \\ 
NGC~6864 &   2.93 &   0.62 & 0.12 & \nodata & \nodata \\ 
NGC~6934 &   2.51 &   0.56 & 0.10 &  0.001&   0.028 \\ 
NGC~6981 &   2.72\tablenotemark{b} &   0.63 & 0.10 &\nodata & \nodata \\ 
~ \\
High Red \\
NGC~6273 &   3.37 &   0.66 & 0.15 &  0.019&   0.023 \\ 
NGC~6284 &   3.05 &   0.70 & 0.13 & \nodata & \nodata \\ 
NGC~6293 &   3.11 &   0.62 & 0.11 & \nodata & \nodata \\ 
NGC~6333 &   3.27 &   0.76 & 0.19 &   0.036&   0.052 \\ 
NGC~6402 &   3.89 &   0.76 & 0.19 &  0.045&   0.064 \\ 
NGC~6440 &   5.68 &   1.04 & 0.31 &  0.100&   0.055 \\ 
NGC~6544 &   4.50 &   0.84 & 0.21 &  0.038&   0.015 \\ 
NGC~6626 &   3.34 &   0.72 & 0.19 &  0.057&   0.055 \\ 
NGC~6638 &   3.83 &   0.77 & 0.15 & \nodata & \nodata \\ 
NGC~6712 &   3.50 &   0.72 & 0.17 &  0.092&   0.064 \\ 
NGC~6779 &   2.92 &   0.58 & 0.13 &  0.033&   0.036 \\ 
~ \\
Others\tablenotemark{d} \\
NGC~288  & 2.11 & 0.60 & 0.13  & \nodata & \nodata \\
NGC~1851 & 2.51 & 0.63 & 0.10 & \nodata & \nodata \\
NGC~2298 & \nodata &  0.59 & 0.13  & \nodata & \nodata \\
NGC~2808 & 2.96 & 0.69 & 0.16 & \nodata & \nodata \\
NGC~4147 & \nodata & 0.47 & 0.12  & \nodata & \nodata \\
NGC~5286 & 2.98 & 0.63 & 0.16  & \nodata & \nodata \\
NGC~5694 & 2.39 & 0.53 & 0.08  & \nodata & \nodata \\
NGC~5824 & 2.53 &  0.59 & 0.12  & \nodata & \nodata \\
NGC~5927 & \nodata & 0.89 & 0.28 &  \nodata & \nodata \\ 
NGC~5986 & 2.93 & 0.66 & 0.17 & \nodata & \nodata \\
NGC~6139 & 4.42 & 0.80 & 0.24 & \nodata & \nodata \\
NGC~6304 &   4.63 & 0.92 & 0.26 & \nodata & \nodata \\ 
NGC~6316 &   4.80 & 0.82 & 0.22 & \nodata & \nodata \\ 
NGC~6342 &   4.05 & 0.81 & 0.22 & \nodata & \nodata \\ 
NGC~6355 &   4.49 & 0.85 & 0.25 & \nodata & \nodata \\ 
NGC~6388 &   3.69 & 0.78 & 0.21 & \nodata & \nodata \\ 
NGC~6441 &   3.90 & 0.83 & 0.22 & \nodata & \nodata \\ 
NGC~6517 &   5.17 & 0.86 & 0.26 & \nodata & \nodata \\ 
NGC~6522 &   3.48 & 0.74 & 0.17 & \nodata & \nodata \\ 
NGC~6528 &   4.52 & 0.89 & 0.23 & \nodata & \nodata \\ 
NGC~6535 &   3.36 & 0.89 & 0.11 & \nodata & \nodata \\ 
NGC~6539 &   5.45 & 1.00 & 0.28 & \nodata & \nodata \\ 
NGC~6553 &   5.51 & 1.04 & 0.32 & \nodata & \nodata \\ 
NGC~6624 &   3.61 & 0.78 & 0.21 & \nodata & \nodata \\ 
NGC~6642 &   3.40 & 0.74 & 0.16 & \nodata & \nodata \\ 
NGC~6681 &   2.52 & 0.61 & 0.12 & \nodata & \nodata \\ 
NGC~6749 &   6.97 & 1.03 & 0.46 & \nodata & \nodata \\ 
NGC~6760 &   4.98 & 0.95 & 0.28 & \nodata & \nodata \\   
\enddata
\tablenotetext{a}{Observed colors of Aaronson, Malkan \& Kleinmann, about
1977, unpublished, see brief description in \cite{acmm}.  These are in the
CIT system, not 
the 2MASS system.  Part of this dataset, in the form of
reddening corrected broad band colors, was published by \cite{brodie90}.} 
\tablenotetext{b}{Discrepancy in $V-K$ 
between the two independent archives of the 1977 data exceeds 0.2 mag.
See text for details.} 
\tablenotetext{c}{$V-K$ not used by \cite{acmm}.}
\tablenotetext{d}{None of these were used by \cite{acmm}.}
\end{deluxetable}

\clearpage

%
%
\begin{deluxetable}{lrrr}
\tablenum{6}
\tablewidth{0pt}
\tablecaption{Comparison of IR Integrated Light Photometry of
Galactic GCs  
\label{table_acmm_comp}}
\tablehead{
\colhead{Group} & \colhead{Number in Common} & 
\colhead{Mean $\Delta$} & \colhead{${\sigma}[$[Mean $\Delta$])} \\
\colhead{} & \colhead{} & \colhead{(mag)} & \colhead{(mag)} \\
\colhead{} & \colhead{} & 
\multispan{2}{2MASS(2006) -- AMK(1978)\tablenotemark{a}} \\
 }
\startdata 
$V-K_s$ \\
Calibrators & 12 & $-0.15$ & 0.25 \\
All         & 35 & $-0.07$ & 0.26 \\
$E(B-V)<0.40$~mag  & 27 & $-0.12$ & 0.25 \\
~~~ \\
$J-K_s$ \\
Calibrators & 12 & $-0.14$ & 0.08 \\
All         & 35 & $-0.13$ & 0.07 \\
$E(B-V)< 1.0)$~mag & 34 & $-0.13$ & 0.07 \\
\enddata
\tablenotetext{a}{Aaronson, Malkan \& Kleinmann (1978), unpublished,
see brief description in \cite{acmm}.}
\end{deluxetable}

%
%
\begin{deluxetable}{lrrr}
\tablenum{7}
\tablewidth{0pt}
\tablecaption{Comparison of IR Integrated Light Photometry of
Galactic GCs From 2MASS -- Us vs. \cite{nantais06}  
\label{table_huchra_comp}}
\tablehead{
\colhead{Group} & \colhead{Number in Common} & 
\colhead{Mean $\Delta$} & \colhead{${\sigma}[$[Mean $\Delta$])} \\
\colhead{} & \colhead{} & \colhead{(mag)} & \colhead{(mag)} \\
\colhead{} & \colhead{} & \multispan{2}{2MASS(us) -- \cite{nantais06}} \\
 }
\startdata 
$V-K_s$ \\
All ``reliable''\tablenotemark{a} & 58 & $+0.63$ & 0.52 \\
~~~ \\
$J-K_s$ \\
All ``reliable''\tablenotemark{a} & 59 & $-$0.01 & 0.12 \\
~~~ \\
$J-H_s$ \\
All ``reliable''\tablenotemark{a} & 59 & $-$0.02 & 0.13 \\
\enddata
\tablenotetext{a}{Includes only those GCs regarded as ``reliable'' by
\cite{nantais06}.  Those with $V-J < 0$, among others, are excluded.}
\end{deluxetable}

\clearpage

%
%
\begin{deluxetable}{lrrrr rr}
\tablenum{8}
\tablewidth{0pt}
\tablecaption{Fits to Integrated Light Colors for Galactic GCs
As a Function of [Fe/H]
\label{table_feh_fit}}
\tablehead{
\colhead{Group} & \colhead{Number GCs} & 
\colhead{Order of Fit\tablenotemark{a}} & \colhead{$A(0)$} &
\colhead{$A(1)$} & \colhead{$A(2)$} &
\colhead{$\sigma$ About Fit} \\
\colhead{} & \colhead{} & \colhead{} & \colhead{(mag)} &
\colhead{} & \colhead{(mag$^{-1}$)}  & \colhead{(mag)} \\
 }
\startdata 
$V-J$ \\
Best &   38  &  1 & 2.15 & 0.268 &  \nodata & 0.17 \\
Best &   38  &  2 & 2.33 & 0.552 & 0.010 & 0.17 \\
Fair &   53  &  1 & 2.15 & 0.282 & \nodata & 0.23 \\
Fair &   53  &  2 & 2.51 & 0.851 & 0.202 & 0.22 \\
Fair + High [Fe/H]\tablenotemark{b} &
        56   &  2 & 2.89 & 1.399 & 0.380 & 0.24\tablenotemark{c} \\
~~~ \\
$V-H$ \\
Best &   38  &  1 & 2.78 & 0.373 &  \nodata & 0.17 \\
Best &   38  &  2 & 3.10 & 0.872 & 0.172 & 0.17 \\
Fair &   53  &  1 & 2.75 & 0.374 & \nodata & 0.24 \\
Fair &   53  &  2 & 3.19 & 1.070 & 0.247 & 0.23 \\
Fair + High [Fe/H]\tablenotemark{b} &
        56   &  2 & 3.57 & 1.610 & 0.423 & 0.24\tablenotemark{c} \\
~~~ \\
$V-K_s$ \\
Best &   38  &  1 & 2.93 & 0.409 &  \nodata & 0.18 \\
Best &   38  &  2 & 3.25 & 0.903 & 0.170 & 0.17 \\
Fair &   53  &  1 & 2.89 & 0.404 & \nodata & 0.24 \\
Fair &   53  &  2 & 3.30 & 1.071 & 0.125 & 0.24 \\
Fair + High [Fe/H]\tablenotemark{b} &
        56   &  2 & 3.59 & 1.481 & 0.371 & 0.23\tablenotemark{c} \\
ACMM(1978)\tablenotemark{d} & 14 & 1 & 2.97$\pm0.11$ & 0.50$\pm0.07$ & \nodata & \nodata \\        
~~~ \\
$J-K_s$ \\
Best &   52  &  1 & 0.785 & 0.135 &  \nodata & 0.06 \\
Best &   52  &  2 & 0.824 & 0.216 & 0.035 & 0.06 \\
Fair &   82  &  1 & 0.794 & 0.148 & \nodata & 0.07 \\
Fair &   82  &  2 & 0.829 & 0.227 & 0.034 & 0.07 \\
Fair + High [Fe/H]\tablenotemark{b} &
        84   &  2 & 0.827 & 0.224 & 0.033 & 0.07 \\
ACMM(1978)\tablenotemark{d} & 14 & 1 & 0.82$\pm0.03$ & 0.14$\pm0.02$  & \nodata & \nodata \\
~~~ \\
$J-H$ \\
Best &   52  &  1 & 0.638 & 0.094 &  \nodata & 0.05 \\
Best &   52  &  2 & 0.673 & 0.167 & 0.030 & 0.05 \\
Fair &   82  &  1 & 0.646 & 0.104 & \nodata & 0.06 \\
Fair &   82  &  2 & 0.669 & 0.155 & 0.022 & 0.06 \\
Fair + High [Fe/H]\tablenotemark{b} &
        84   &  2 & 0.672 & 0.159 & 0.023 & 0.06 \\
\enddata
\tablenotetext{a}{Fit is linear (1) or quadratic (2)}
\tablenotetext{b}{Adds those GCs in our sample with 
[Fe/H] $> -0.2$ dex that are not already included.  See text for
details.} 
\tablenotetext{c}{$\sigma$ rises to $\sim$0.30 mag if a linear fit is used.}
\tablenotetext{d}{1977 fit transformed from Johnson to 2MASS colors.}
\end{deluxetable}

\clearpage

\end{document}